\documentclass[twocolumn]{emulateapj}
\usepackage{graphicx}
\usepackage{dcolumn}

\begin{document}

\title{ The Link Between Star Formation and Accretion in LINERs: A Comparison with other AGN Subclasses}

\author{S. Satyapal, \altaffilmark{1,2} R. P. Dudik, \altaffilmark{1}  B. O'Halloran, \altaffilmark{1}  \& M. Gliozzi\altaffilmark{1}}

\altaffiltext{1}{George Mason University, Department of Physics \& Astronomy, MS 3F3, 4400 University Drive, Fairfax, VA 22030}

\altaffiltext{2}{Presidential Early Career Award Scientist}

\begin{abstract}

We present archival high resolution X-ray imaging observations of 25 nearby LINERs observed by ACIS on board {\it Chandra}.  This sample builds on our previously published proprietary and archival X-ray observations, and includes the complete set of LINERs with published black hole masses and FIR luminosities that have been observed by {\it Chandra}.  Of the 82 LINERs observed by {\it Chandra}, 41 (50\%)  display hard nuclear cores consistent with an AGN. The nuclear 2-10 keV luminosities of these AGN-LINERs range from $\sim$ 2 $\times$ 10$^{38}$ ergs s$^{-1}$ to $\sim$ 1 $\times$ 10$^{44}$ ergs s$^{-1}$.  Reinforcing our previous work, we find a significant correlation between the Eddington ratio, L$_{\rm bol}$/L$_{\rm Edd}$, and the far-IR (FIR) luminosity, L$_{\rm FIR}$, as well as the IR brightness ratio, L$_{\rm FIR}$/L$_{\rm B}$ in the host galaxy of AGN-LINERs that extends over seven orders of magnitude in L$_{\rm bol}$/L$_{\rm Edd}$.  Combining our AGN-LINER sample with galaxies from other AGN subclasses, we find that this correlation is reinforced in the full sample of 129 AGN, extending over almost nine orders of magnitude in L$_{\rm bol}$/L$_{\rm Edd}$. Using archival and previously published observations of the 6.2 $\mu$m PAH feature from the Infrared Space Observatory ({\it ISO}), we find that it is unlikely that dust heating by the AGN dominates the FIR luminosity in our sample of AGN.  Our results may therefore imply a fundamental link between the mass accretion rate ($\dot{M}$), as measured by the Eddington ratio, and the star formation rate (SFR), as measured by the FIR luminosity. Apart from the overall correlation, we find that the different AGN subclasses occupy distinct regions in the L$_{\rm FIR}$  and L$_{\rm bol}$/L$_{\rm Edd}$ plane.  Assuming a constant radiative efficiency for accretion, our results may imply a variation in the SFR/$\dot{M}$ ratio as function of AGN activity level--a result that may have significant consequences on our understanding of galaxy formation and black hole growth. 

\end{abstract}


\section{Introduction}

With the recent discovery that virtually all local galaxies harbor massive nuclear black holes, there is now convincing evidence that active galactic nuclei (AGN) and normal galaxies in our local Universe are fundamentally connected. However, the nature of this connection and the detailed evolutionary history connecting these objects is unknown. Low Ionization Nuclear Emission Line Regions (LINERs), defined by their narrow optical emission lines of low ionizatation uncharacteristic of photoionization by normal main--sequence stars (Heckman 1980), may constitute a vital piece of this puzzle, possibly representing the ``{\em missing link}'' between the powerful AGN in the Universe and galaxies such as our own. Indeed, the low mass accretion rates inferred for many accretion-powered LINERs (``{\em AGN-LINERs}''; e.g., Ho 1999, Eracleous et al. 2002)  suggest that these objects represent the population of AGN just before accretion onto the black hole ceases. Despite several decades of intense research (e.g. Heckman 1980; Ho et al. 1997a,b; Terashima \& Wilson 2002 and references therein) there are still open questions, including: what fraction of LINERs are truly AGN, what are their accretion properties, and how do these quantities relate to the properties of the host galaxy?

     In an effort to determine the fraction of LINERs hosting AGN and to characterize their accretion properties, we undertook a {\it Chandra} snapshot survey of the largely unexplored population of nearby infrared(IR)-bright LINERs (L$_{\rm FIR}$/L$_{\rm B}$ $>$ 3, D $<$ 30 Mpc) to search for hard nuclear point sources indicative of an AGN (Satyapal, Sambruna, \& Dudik 2004, Dudik et al. 2004; henceforth Papers I \& II).  Combining these observations with archival data of other LINERs with a large range of IR luminosities, we found a correlation between the Eddington ratio, L$_{\rm bol}$/L$_{\rm Edd}$, and the far-IR (FIR) luminosity, L$_{\rm FIR}$ as well as the IR brightness ratio, L$_{\rm FIR}$/L$_{\rm B}$, in the host galaxy of AGN-LINERs that extends over seven orders of magnitude in L$_{\rm bol}$/L$_{\rm Edd}$.  This trend implies that either the mass accretion rate or the radiative efficiency, or a combination of both, scales with the FIR luminosity and IR-brightness ratio of the host galaxy.  If the FIR luminosity is associated primarily with star formation, this correlation may imply a fundamental link between accretion onto the black hole and the star formation rate (SFR)/age of the host galaxy. If the FIR luminosity is also an indicator of the molecular gas content in this sample of LINERs, our results may further indicate that the mass accretion rate scales with the host galaxy's fuel supply (Paper II).

	A correlation between the mass accretion rate and the properties of current star formation in LINERs may have profound implications for our understanding of black hole growth and the connection between starbursts and AGN.  The well-known tight correlation between black hole mass and the stellar velocity dispersion (Gebhardt et al. 2000, Ferrarese \& Merritt 2000) implies that black hole and bulge formation and growth are intimately connected.  This relationship appears to hold for active galaxies (e.g. Gebhardt et al. 2000b, Ferrarese et al. 2001, Nelson 2000, Wandel 2002, Boroson 2003, and Shields et al. 2003), although there is some controversy over whether it applies to all AGN classes (e.g., Grupe \& Mathur 2004, Bian \& Zhao 2004, Botte et al. 2004).  These results imply an intimate relationship between the formation and growth of black holes and the surrounding galactic bulge even when the black hole is actively accreting.  By studying a potentially direct tracer of current star formation such as the FIR luminosity, we can explore the connection between star formation and black hole growth in various AGN subclasses.

Since the FIR luminosity in the galaxies presented in Paper II is measured in a large aperture that includes the emission from the entire host galaxy, the interpretation of the correlations presented in Paper II is uncertain.  Although LINERs tend to harbor weak AGN, it is not clear if the FIR luminosity in these objects is produced primarily by star formation.  The FIR emission can include thermally reprocessed radiation from the AGN, and, additionally, the fraction of the total FIR luminosity attributable to the AGN may increase with AGN power, as measured by the Eddington ratio.  In such cases, observations of the mid-IR (MIR) Polycyclic Aromatic Hydrocarbon (PAH) dust features are a powerful tool for identifying regions of recent star formation and assessing their contribution to the total FIR emission.  These features show a strong correlation with the FIR luminosity in galaxies dominated by starbursts but are absent or weak in galaxies dominated by AGN (e.g., Genzel et al. 1998, Rigopoulou et al. 1999, Clavel et al. 2000).   Consequently, we present in this work, archival and previously published observations of the of the 6.2 $\mu$m emission feature, one of the most widely-used MIR tracers of star formation (e.g. Peeters, Spoon, \& Tielens 2004 and references therein), of all objects presented in this paper that were observed by the {\it Infrared Space Observatory  (ISO)}\footnote[1]{See www.iso.vilspa.esa.es/science/pub\_credit.html}.  These observations are used to determine whether the correlation between L$_{\rm bol}$/L$_{\rm Edd}$ and L$_{\rm FIR}$ indeed implies a connection between the  SFR and the growth of the black hole. 

   The correlation between L$_{\rm bol}$/L$_{\rm Edd}$ and L$_{\rm FIR}$, L$_{\rm FIR}$/L$_{\rm B}$ presented in Paper II was based on a limited sample of 20 LINERs with confirmed AGN in their nucleus.   In an effort to confirm and assess the statistical significance of this correlation, in this paper we present a study of the 25 remaining LINERs in the {\it Chandra} archive for which stellar velocity dispersion-derived black hole masses are available. Of these LINERs, 13 show hard nuclear point sources consistent with an AGN. In addition, to assess whether the correlation is unique to LINERs or whether it extends to other AGN subclasses, in this paper we combine our data with published FIR measurements, bolometric luminosity and black hole mass estimates of additional galaxies that belong to other AGN subclasses. 
   
   This paper is organized as follows.  In Section 2, we summarize the properties of the new {\it Chandra} archival sample presented in this paper, along with a description of additional samples of the various AGN subclasses included in our analysis.  In Section 3, we summarize the observations and data analysis procedure employed for both our {\it Chandra} and {\it ISO} observations, followed by a description of our results, including a discussion of our correlations for the various AGN subclasses in Section 4.  In Section 5, we discuss the implications of our correlations, followed by a summary of our major conclusions in Section 6. 

\section{The Sample}
In this paper, we expand our previously published sample of LINERs (Papers I and II) and include in our analysis data from other AGN subclasses in the literature.  In order to include additional AGN, we must be able to estimate the mass of the black hole and the bolometric luminosity of the AGN.  Black hole masses in AGN are derived primarily through: 1) resolved stellar kinematics, 2) reverberation mapping, or 3) applying the correlation between optical bulge luminosity and central black hole mass determined in nearby galaxies. Dynamical estimates of the central mass are the most reliable but they exist for only a handful of AGN (e.g., see review by Ferrarese \& Ford 1999).  In the absence of dynamical estimates, black hole mass estimates based on reverberation mapping observations are generally accepted to be accurate to within a factor of 3 (e.g., Peterson et al. 2004).  Since the optical light from the bulge is often overwhelmed by that from the AGN, black hole mass estimates based on large aperture measurements of the optical luminosity are less reliable. In this work, we conservatively included a selection of AGN for which the black hole mass was derived either through dynamical estimates, reverberation mapping, or host galaxy bulge luminosity {\it only} if the host galaxy was clearly resolved.  In addition, in order to be included in our analysis, IRAS FIR fluxes and reliable bolometric luminosities must be available for all galaxies. Since the number of such objects is limited, we emphasize that these samples are not complete in any sense and are therefore subject to selection biases.  Each sample is discussed separately below.  The assumptions employed in estimating black hole masses and bolometric luminosities are discussed in each case.

\subsection{LINER Sample}

We searched the {\it Chandra} archive for all galaxies classified as LINERs in the multifrequency LINER catalog from Carillo et al. (1999) which were not previously published in our earlier works (Paper I; Paper II) \footnote[2]{We include all galaxies that are classified as LINERs using {\it either} the Heckman (1980) or the Veilleux \& Osterbrock (1987) diagnostic diagrams.}. Of the 43 newly available galaxies, only 25 had stellar velocity dispersions and FIR fluxes published in the literature.  These galaxies are nearby (1 to 215 Mpc; d$_{\rm median}$=13 Mpc), and comprise a wide range of FIR luminosities (5.87 $\times$ 10$^{6}$ L$_{\odot}$ to 3.26$\times$ 10$^{10}$ L$_{\odot}$, infrared-to-blue ratios (0.01 to 2.81), and Hubble types.  We combine this sample with the compilation presented in Papers I and II.  The entire {\it Chandra} sample consists of a total of 82 galaxies. In this paper we refer to the combined sample as the {\it ``comprehensive LINER sample''}, and the sample from this paper alone as the {\it ``archival LINER sample''}.  The basic properties of the archival LINER sample are summarized in Table 1 and Figure 1.  Following the treatment described in Papers I and II (see Section 3), LINERs that displayed a dominant hard nuclear point source are classified as AGN-LINERs.  Black hole masses for these sources were determined using the correlation between the stellar velocity dispersion and black hole mass (see Table 1; Ferrarese \& Merritt 2000; Gebhardt et al. 2000) demonstrated to hold for nearby AGN (Ferrarese et al. 2001; McLure \& Dunlop 2002).  While all of the LINERs in the archival LINER sample possess black hole mass estimates, seven AGN-LINERs from Paper II do not have this data.  They have therefore been excluded from those plots requiring such estimates.  We note that of the 25 LINERs in the archival LINER sample, 13 display hard nuclear point sources consistent with an AGN and are used in the main study of this work.  We refer to these AGN as the {\it ``archival AGN-LINER sample''}.  As in Paper II, the bolometric luminosity of all targets was calculated using the formula L$_{\rm bol}$ = 34$\times$ L$_{\rm X}$(2-10keV), determined from the spectral energy distribution(SED) measurements of seven low luminosity AGN from Ho 1999 (see Paper II for details).

 Of the 82 LINERs in our comprehensive LINER sample, 41 are AGN-LINERs, 34 of which have published black hole masses.  Since AGN-LINERs with published black hole masses are the main study of this paper, we refer to the sample of all AGN-LINERs with published black hole masses included in this work as the {\it ``comprehensive AGN-LINER sample''}. The galaxy properties were taken from the NASA/IPAC Extragalactic Database (NED). We note that the comprehensive LINER sample published in this work includes the complete set of LINERs from the Carillo et al. (1999) catalog with published black hole mass estimates that exist in the {\it Chandra} archive.

\begin{figure*}[]
{\includegraphics[width=8cm]{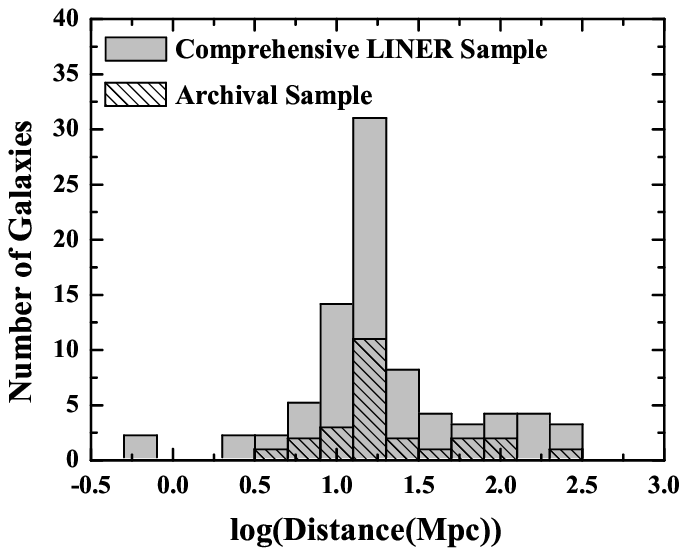}}{\includegraphics[width=8cm]{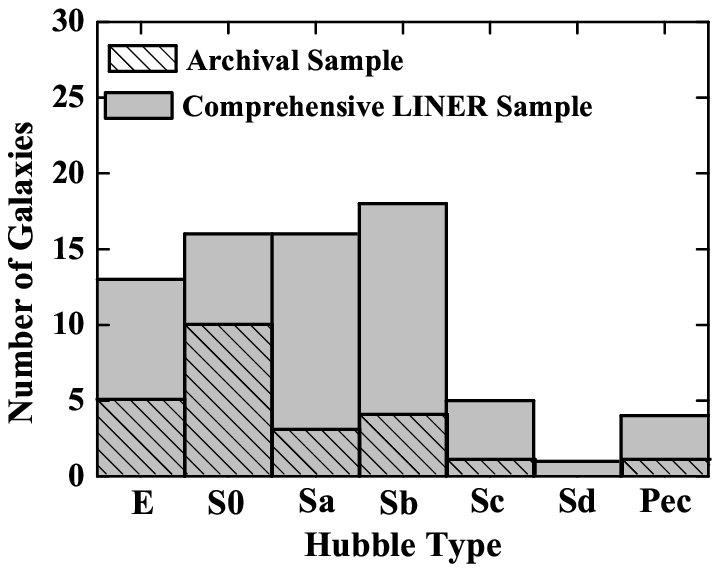}}\\
{\includegraphics[width=8cm]{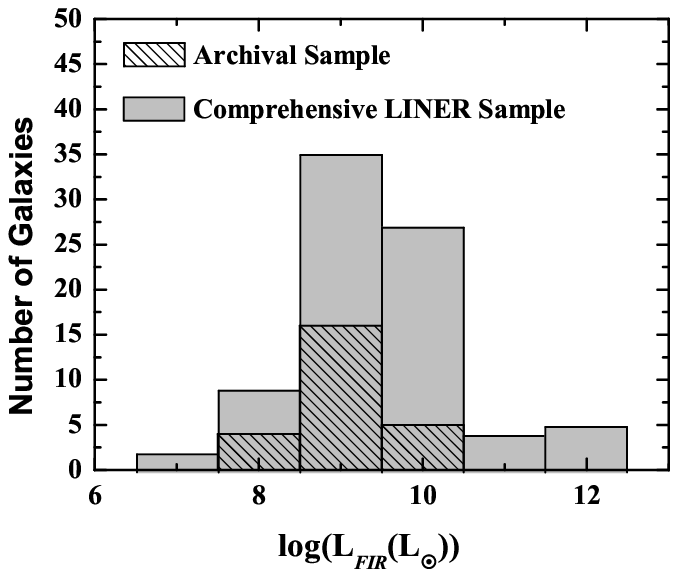}}{\includegraphics[width=8cm]{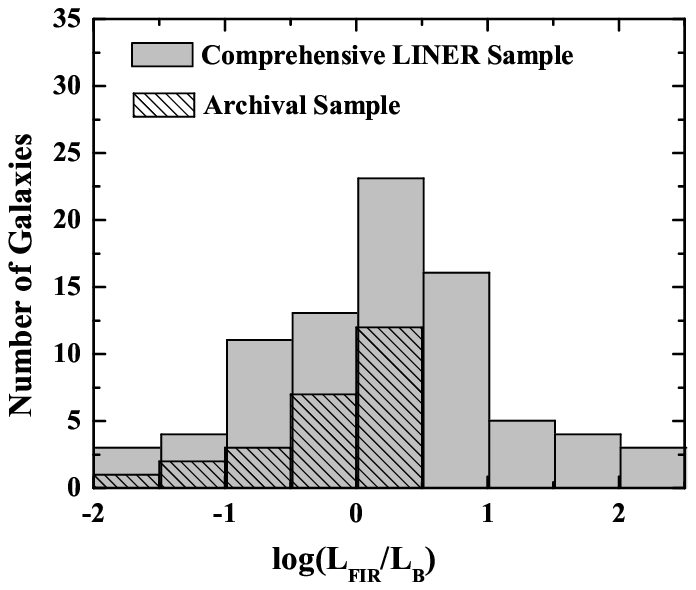}}\\
\caption[]{Characteristics of the comprehensive {\it Chandra} sample of LINERs from this paper and Paper II.  The solid bins correspond  to the entire sample from this work and Paper II (the {\it ``comprehensive LINER sample''}). The striped bins correspond to the LINERs from this paper only (the {\it ``archival LINER sample''}).  Most galaxies in the comprehensive LINER sample are nearby and span a wide range of luminosities, IR-brightness ratios, and Hubble types. Note that not all galaxies have Hubble classifications available in the literature and have therefore been excluded from that plot.}
\end{figure*}

\begin{table*}
\begin{center}
\begin{tabular}{lccccc}
\multicolumn{6}{l}{{\bf Table 1: Properties of the Archival Sample}}  \\
 \hline\hline
\multicolumn{1}{c}{Galaxy} & \multicolumn{1}{c}{Distance} & Hubble & log & log 
& log \\
\multicolumn{1}{c}{Name} & \multicolumn{1}{c}{(Mpc)} & Type &
L$_{\rm FIR}$ & (L$_{\rm FIR}$/L$_{\rm B}$) &  (M$_{\rm BH}$)  \\

\multicolumn{1}{c}{(1)} & \multicolumn{1}{c}{(2)} & (3) & (4) & (5) & (6) \\
\hline\hline
\multicolumn{6}{l}{{\it AGN-LINERs}}\\
\hline
NGC0315 & 66 & E & 9.4 & -0.69 & 9.12 \\
NGC2681 & 13$^*$ & SAB(rs)0 & 8.8 & -0.30 & 7.75 \\
NGC3169 & 20$^*$ & SA(s)a & 9.8 & 0.30 & 7.86 \\
NGC3245 & 22$^*$ & SA(r)0 & 9.2 & -0.08 & 8.39 \\
NGC3718 & 17$^*$ & SB(s)a & 8.7 & -0.46 & 7.93 \\
NGC4258 & 7$^*$ & SAB(s)bc & 9.4 & 0.06 & 7.16 \\
NGC4261 & 35$^*$ & E2-3 & 8.9 & -0.96 & 8.84 \\
NGC4410A & 97 & Sab? & 9.7 & 0.001 & 8.73 \\
NGC4457 & 17$^*$ & SAB(s)0 & 9.4 & 0.35 & 6.95 \\
NGC4552 & 17$^*$ & E & 8.0 & -1.44 & 8.57 \\
NGC4565 & 10$^*$ & SA(s)b? & 9.2 & 0.14 & 7.56 \\
NGC6482 & 52$^*$ & E & 8.8 & -1.01 & 8.92 \\
3C218 & 215 & (R')SA0 & 10.2 & -0.23 & 8.88 \\
 \hline\hline
\multicolumn{6}{l}{\it Non-AGN-LINERs}\\
\hline
NGC2541 & 11$^*$ & SA(s)cd & 8.5 & 0.08 & 5.81 \\
NGC2683 & 6$^*$ & SA(rs)b & 8.5 & -0.01 & 7.36 \\
NGC4410B & 97 & S0 & 9.8 & 0.45 & 8.23 \\
NGC4150 & 10$^*$ & SA(r)0 & 8.3 & 0.01 & 7.36 \\
NGC4438 & 17$^*$ & SA(s)0 & 9.4 & 0.02 & 5.19\\
NGC4459 & 17$^*$ & SA(r)0 & 9.0 & -0.19 & 7.85 \\
NGC4501 & 17$^*$ & SA(rs)b & 10.0 & 0.39 & 7.64 \\
NGC4548 & 17$^*$ & SBb(rs) & 9.2 & -0.20 & 7.68 \\
NGC4550 & 17$^*$ & SB0 & 7.8 & -0.92 & 6.79 \\
NGC4736 & 4$^*$ & SA(r)ab & 9.3 & 0.28 & 7.07 \\
NGC5846 & 29$^*$ & E0-1 & 7.8 & -1.94 & 8.59 \\
NGC5866 & 15$^*$ & S03 & 9.4 & 0.05 & 7.49 \\
 \hline\hline
\end{tabular}
\end{center}
{\scriptsize{\bf Columns Explanation:} Col(1):Common Source Names; Col(2):  Distance.  Since most of these galaxies are nearby, we have taken distances (marked $^*$) from Tully 1988 who derived distances based on the Virgo infall model.  All others were calculated using redshift for H$_0$= 75 km s$^{-1}$Mpc$^{-1}$; Col(3): Morphological Class; Col(4): Far-infrared luminosities (in units of solar luminosities: L$\odot$) correspond to the 40-500$\mu$m wavelength interval and were calculated using the IRAS 60 and 100 $\mu$m fluxes according to the prescription: $L_{FIR}$=1.26$\times$10$^{-14}$(2.58f$_{60}$+f$_{100}$) in W m$^{-2}$ (Sanders \& Mirabel 1996).; Col(5):  L$_{B}$: B magnitude see Carrillo et al (1999); Col(6):  Mass of central black hole calculated using the stellar velocity despersion in the formula:  M$_{\rm BH}$ =  1.2($\pm$0.2)$\times$10$^8$ M$_{\odot}$($\sigma$$_e$/200 km s$^{-1}$)$^{3.75 ({\pm} 0.3)}$)  (from Ferrarese \& Merritt 2000; Gebhardt et al. 2000a; Tremaine et al. 2002). Velocity dispersions are taken from the Hypercat database available online at http://www-obs.univ-lyon1.fr/hypercat.}
\end{table*}

\subsection{Seyfert Sample}
 The Seyfert galaxies included in our analysis were taken from the sample compiled by Woo \& Urry (2002).  Six of the galaxies in this sample were excluded because they are cross-listed in other AGN-subclass samples we include in this work (2 are Narrow Line Seyfert1s, 2 are LINERs, and 2 are radio loud AGN).  Only those objects with both 60 and 100 $\mu$m and blue magnitudes available in NED were selected.  These include 12 Seyfert 1 galaxies with black hole masses derived from reverberation mapping (Kaspi et al. 2000, or Onken \& Peterson 2002), 2 Seyfert 1s with M$_{\rm BH}$ derived through their optical luminosity, and the remaining Seyfert 1s and 2s with M$_{\rm BH}$ derived from stellar velocity dispersion measurements.  All Seyferts are nearby, spanning a distance range similar to that spanned by our LINER sample (d=4 to 136 Mpc; d$_{\rm median}$ = 56 Mpc).  Bolometric luminosities were determined in virtually all cases by integrating all available flux points in the well-sampled SED.  We note that since the two galaxies with black hole mass estimates derived from their optical luminosity are nearby, spatially resolved observations of their nuclear optical luminosity were possible.

\subsection{Radio-Quiet Quasar Sample}
The radio-quiet quasars (RQQ) included in our analysis were also taken from Woo \& Urry (2002).  Again, only those objects with firm 60 and 100 $\mu$m IRAS detections and blue magnitudes available in NED were selected.  Ten out of 15 of these objects have M$_{\rm BH}$ derived from reverberation mapping, and the remaining 5 have M$_{\rm BH}$ derived through their optical luminosity.  Bolometric luminosities were determined through either direct flux integration of the SED or by flux-fitting the appropriate RQQ template SED (Elvis et al. 1994) to the available flux points.  The RQQ sample spans a distance range from z=0.06 to z=0.29, with a median z=0.11.

\subsection{Radio-Loud AGN Sample}
In radio-loud AGN, the stellar light is often overwhelmed by the nonthermal contribution from the AGN.  As a result, black hole mass estimates based on the host galaxy's bulge luminosity are unreliable.  We therefore use the sample of radio-loud AGN from Marchesini et al. (2004), which includes only those objects in the complete sample of 53 RLQs in the 3CR catalog (Spinrad et al. 1985) with clearly resolved host galaxies.  We have included only those objects with firm IRAS detections and blue magnitudes, and of these 17, three additional galaxies have been excluded because they exist in our comprehensive AGN-LINER sample.  The final radio-loud sample comprises 4 radio galaxies exhibiting Fanaroff-Riley type I (FR I) radio morphology, 5 objects exhibiting FR II morphology (Fanaroff \& Riley 1974), and 5 radio-loud quasars.  The sample spans a distance range from z=0.017 to z=1.436, with a median distance of z=0.07.  Black hole masses were derived either through stellar velocity dispersion measurements or through the host galaxy's optical luminosity. Bolometric luminosities were obtained from the rest-frame monochromatic luminosity at 5100$\AA$  (McLure \& Dunlop 2001) using the bolometric correction from Elvis et al. (1994).

We note that several LINERs are radio-loud (e.g. Ho 1999; Terashima et al. 2003) and are found to exhibit weak radio jets (e.g. Yuan et al. 2002, Nagar et al. 2004).  Alternatively, several radio galaxies that display either FR I or FR II radio morphologies also exhibit weak optical emission lines with LINER-like line ratios (e.g Tadhunter et al. 1993, Lewis et al. 2003).  Strictly speaking, our LINER sample and the radio-loud AGN sample may not therefore be distinct AGN subclasses. However, in order to assess the incidence of possible selection biases, to enlarge the statistics, and to include radio-loud quasars (RLQs) with redshifts overlapping with the redshift range of the RQQs, we include this sample in our analysis. 

\subsection{Narrow Line Seyfert 1 Sample}
A subset of Seyfert galaxies display narrow permitted optical lines (NLS1s; Osterbrock \& Pogge 1985), appear to accrete at close to the Eddington rate and have smaller mass black holes for a  given luminosity compared to regular Seyfert 1s (e.g. Borosson 2002, Grupe et al. 2004).  In order to expand our sample to include low values of M$_{\rm BH}$ and high L$_{\rm bol}$/L$_{\rm Edd}$ values, we include the sample of NLS1s from Grupe et al. 2004, which consists of a complete sample of 110 soft X-ray selected AGNs.  Of the 51 NLS1s in this sample, only 12 have 60 and 100 $\mu$m IRAS detections and 4 more have 100 $\mu$m IRAS upper limits.  Of these 16 galaxies, 2 were excluded because they overlap with our RQQ sample.  Black hole masses were calculated using the H$\beta$ line width and 5100~$\AA$ luminosity using the empirical relation from Kaspi et al. (2000). Bolometric luminosities were estimated by using a combined power-law model fit with exponential cutoff to the optical-UV data (See Grupe et al. 2004 for details). Our final NLS1 sample consists of 14 objects with distances that range from z=0.02 to z=0.14, with a median distance of z=0.045.

The entire AGN sample included in our analysis, which we refer to as {\it ``The Expanded AGN sample''},  consists of: 52 Seyfert galaxies, 14 radio-loud AGN, 15 RQQ, and 14 NLS1.  Combined with our comprehensive AGN-LINER sample, the total number of galaxies in our analysis is 129.  We emphasize again that the basis for selection of objects in the expanded AGN sample is on the availability of reliable black hole mass, bolometric and FIR luminosities.  The sample therefore should not be viewed as complete in any sense.  Black hole masses, bolometric luminosities, and Eddington ratios for all objects included in our Expanded AGN sample are listed in Table 2. 

\begin{table*}
\begin{center}
\begin{tabular}{lcccccc}
\multicolumn{7}{l}{\bf Table 2: Expanded AGN Sample Properties}  \\ 
 \hline\hline
\multicolumn{7}{l}{\bf {\it 2a: AGN-LINERs from Papers I and II$^+$}}  \\ 
 \hline

\multicolumn{1}{c}{Galaxy Name} & \multicolumn{1}{c}{{\it z}} & 
log(M$_{\rm BH}$)$^a$ & log(L$_{\rm bol}$)$^a$ & log(L$_{\rm FIR}$) & log( L$_{\rm FIR}$/L$_{\rm B}$) & log(L$_{\rm bol}$/L$_{\rm Edd}$)\\

\multicolumn{1}{c}{(1)} & \multicolumn{1}{c}{(2)} & (3) & (4) & (5) & (6) & 
(7)\\ 
\hline\hline
NGC3125 & 0.0029 & 5.77 & 41.5 & 9.0 & 0.95 & -2.39 \\
NGC 4350 & 0.0040 & 7.99 & 40.4 & 8.3 & -0.70 & -5.75 \\
NGC1052 & 0.0049 & 8.29 & 43.2 & 9.1 & -0.56 & -3.24 \\
NGC3031 & -0.0001 & 7.79 & 41.7 & 8.4 & -0.85 & -4.16 \\
NGC4278 & 0.0022 & 9.20 & 41.6 & 8.5 & -0.75 & -5.68 \\
NGC4486 & 0.0044 & 9.48 & 42.1 & 8.3 & -1.57 & -5.53 \\
NGC4579 & 0.0051 & 7.85 & 42.5 & 9.5 & -0.03 & -3.47 \\
NGC6500 & 0.0100 & 8.82 & 41.8 & 9.4 & 0.14 & -5.16 \\
NGC 4203 & 0.0036 & 7.90 & 41.6 & 8.5 & -0.42 & -4.39 \\
NGC 4494 & 0.0045 & 7.65 & 40.4 & 7.7 & -1.73 & -5.36 \\
NGC 4594 & 0.0034 & 9.04 & 41.7 & 9.0 & -0.68 & -5.47 \\ 
NGC4527 & 0.0058 & 8.25 & 40.3 & 9.9 & 0.94 & -5.54 \\
NGC4125 & 0.0045 & 8.50 & 40.2 & 8.6 & -0.96 & -6.44 \\
NGC4374 & 0.0035 & 9.20 & 40.6 & 8.5 & -0.51 & -6.65 \\
NGC4696 & 0.0099 & 8.60 & 41.7 & 8.8 & -1.11 & -5.05 \\
NGC5194 & 0.0015 & 6.90 & 42.6 & 9.8 & 0.23 & -2.43 \\
MRK273 & 0.0378 & 7.74 & 45.6 & 11.8 & 2.21 & -0.28 \\
CGCG162-010 & 0.0633 & 9.07 & 43.4 & 10.6 & 0.55 & -3.79 \\
NGC6240 & 0.0245 & 9.15 & 45.7 & 11.3 & 1.58 & -1.51 \\
IC1459 & 0.0056 & 9.00 & 41.9 & 8.6 & -1.03 & -5.15 \\
NGC 2787 & 0.0023 & 7.59 & 39.8 & 7.8 & -0.54 & -5.85 \\

\hline\hline
\end{tabular}
\end{center}
\end{table*}

\begin{table*}
\begin{center}
\begin{tabular}{lcccccc}
\multicolumn{7}{l}{\bf Table 2: Expanded AGN Sample Properties cont.}  \\ 
 \hline\hline
\multicolumn{7}{l}{\bf {\it 2b: Type 1 Seyferts}}  \\ 
 \hline

\multicolumn{1}{c}{Galaxy Name} & \multicolumn{1}{c}{{\it z}} & 
log(M$_{\rm BH}$)$^a$ & log(L$_{\rm bol}$)$^a$ & log(L$_{\rm FIR}$) & log( L$_{\rm FIR}$/L$_{\rm B}$) & log(L$_{\rm bol}$/L$_{\rm Edd}$)\\

\multicolumn{1}{c}{(1)} & \multicolumn{1}{c}{(2)} & (3) & (4) & (5) & (6) & 
(7)\\ 
\hline\hline

NGC 1566 & 0.005 & 6.92 & 44.5 & 10.2 & 0.30 & -0.57\\
NGC 3227 & 0.004 & 7.64 & 43.9 & 9.6 & 0.34 & -1.88\\
NGC 3516 & 0.009 & 7.36 & 44.3 & 9.6 & 0.13 & -1.17\\
NGC 3783 & 0.010 & 6.94 & 44.4 & 10.0 & 0.47 & -0.63\\
NGC 3982 & 0.004 & 6.09 & 43.5 & 9.6 & 0.57 & -0.65\\
NGC 3998 & 0.003 & 8.95 & 43.5 & 8.2 & -0.65 & -3.51\\
NGC 4151 & 0.003 & 7.13 & 43.7 & 9.2 & 0.27 & -1.50\\
NGC 4593 & 0.009 & 6.91 & 44.1 & 9.9 & 0.10 & -0.92\\
NGC 5548 & 0.017 & 8.03 & 44.8 & 9.9 & 0.26 & -1.30\\
NGC 6814 & 0.005 & 7.28 & 43.9 & 9.8 & 0.64 & -1.46\\
NGC 7469 & 0.016 & 6.84 & 45.3 & 11.3 & 1.51 & 0.34\\
Mrk 10 & 0.029 & 7.47 & 44.6 & 10.4 & 0.24 & -0.96\\
Mrk 79 & 0.022 & 7.86 & 44.6 & 10.3 & 0.65 & -1.39\\
Mrk 509 & 0.034 & 7.86 & 45.0 & 10.6 & -0.27 & -0.93\\
Mrk 590 & 0.026 & 7.20 & 44.6 & 10.1 & 0.27 & -0.67\\
Mrk 817 & 0.032 & 7.60 & 45.0 & 10.7 & 0.98 & -0.71\\
IC 4329A & 0.016 & 6.77 & 44.8 & 10.1 & 0.73 & -0.09\\
UGC 3223 & 0.016 & 7.02 & 44.3 & 10.1 & 0.69 & -0.85\\
Akn 120 & 0.032 & 8.27 & 44.9 & 10.3 & 0.37 & -1.46\\
\hline\hline
\end{tabular}
\end{center}
\end{table*}

\begin{table*}
\begin{center}
\begin{tabular}{lcccccc}
\multicolumn{7}{l}{\bf Table 2: Expanded AGN Sample Properties cont.}  \\ 
 \hline\hline
\multicolumn{7}{l}{\bf {\it 2c: Type 2 Seyferts}}  \\ 
 \hline

\multicolumn{1}{c}{Galaxy Name} & \multicolumn{1}{c}{{\it z}} & 
log(M$_{\rm BH}$)$^a$ & log(L$_{\rm bol}$)$^a$ & log(L$_{\rm FIR}$) & log( L$_{\rm FIR}$/L$_{\rm B}$) & log(L$_{\rm bol}$/L$_{\rm Edd}$)\\

\multicolumn{1}{c}{(1)} & \multicolumn{1}{c}{(2)} & (3) & (4) & (5) & (6) & 
(7)\\ 
\hline\hline

NGC 513 & 0.002 & 7.65 & 42.5 & 8.4 & 0.81 & -3.23\\
NGC 788 & 0.014 & 7.51 & 44.3 & 9.4 & -0.22 & -1.28\\
NGC 1068 & 0.004 & 7.23 & 45.0 & 10.9 & 0.98 & -0.35\\
NGC 1320 & 0.009 & 7.18 & 44.0 & 9.7 & 0.54 & -1.26\\
NGC 1358 & 0.013 & 7.88 & 44.4 & 9.4 & -0.21 & -1.61\\
NGC 1386 & 0.003 & 7.24 & 43.4 & 9.2 & 0.51 & -1.96\\
NGC 1667 & 0.015 & 7.88 & 44.7 & 10.7 & 0.92 & -1.29\\
NGC 2273 & 0.006 & 7.30 & 44.1 & 9.2 & 0.33 & -1.35\\
NGC 3185 & 0.004 & 6.06 & 43.1 & 8.9 & 0.36 & -1.08\\
NGC 4258 & 0.001 & 7.62 & 43.5 & 9.0 & 0.06 & -2.27\\
NGC 5273 & 0.004 & 6.51 & 43.0 & 8.6 & -0.14 & -1.58\\
NGC 5347 & 0.008 & 6.79 & 43.8 & 9.5 & 0.36 & -1.08\\
NGC 6104 & 0.028 & 7.60 & 43.6 & 10.2 & 0.46 & -2.10\\
NGC 7213 & 0.006 & 7.99 & 44.3 & 9.6 & -0.11 & -1.79\\
NGC 7603 & 0.030 & 8.08 & 44.7 & 10.4 & 0.54 & -1.52\\
NGC 7743 & 0.006 & 6.59 & 43.6 & 9.2 & 0.01 & -1.09\\
Mrk 1 & 0.016 & 7.16 & 44.2 & 10.2 & 1.27 & -1.06\\
Mrk 3 & 0.014 & 8.65 & 44.5 & 10.3 & 1.07 & -2.21\\
Mrk 270 & 0.010 & 7.60 & 43.4 & 8.9 & 0.01 & -2.33\\
Mrk 348 & 0.015 & 7.21 & 44.3 & 9.9 & 0.54 & -1.04\\
Mrk 533 & 0.029 & 7.56 & 45.2 & 11.1 & 1.20 & -0.51\\
Mrk 573 & 0.017 & 7.28 & 44.4 & 10.0 & 0.43 & -0.94\\
Mrk 622 & 0.023 & 6.92 & 44.5 & 10.2 & 0.81 & -0.50\\
Mrk 686 & 0.014 & 7.56 & 44.1 & 9.6 & 0.24 & -1.55\\
Mrk 917 & 0.024 & 7.62 & 44.8 & 10.8 & 1.33 & -0.97\\
Mrk 1040 & 0.017 & 7.64 & 44.5 & 10.4 & 0.77 & -1.21\\
Mrk 1066 & 0.012 & 7.01 & 44.6 & 10.6 & 1.36 & -0.56\\
Mrk 1157 & 0.015 & 6.83 & 44.3 & 10.1 & 0.79 & -0.66\\
UGC 3995 & 0.016 & 7.69 & 44.4 & 9.8 & 0.14 & -1.40\\
UGC 6100 & 0.029 & 7.70 & 44.5 & 10.2 & 0.49 & -1.32\\
IC 5063 & 0.011 & 7.74 & 44.5 & 10.2 & 0.77 & -1.31\\
F 341 & 0.016 & 7.15 & 44.1 & 9.8 & 0.54 & -1.12\\
II ZW55 & 0.025 & 8.23 & 44.5 & 10.7 & 1.25 & -1.79\\
\hline\hline
\end{tabular}
\end{center}
\end{table*}

\begin{table*}
\begin{center}
\begin{tabular}{lcccccc}
\multicolumn{7}{l}{\bf Table 2: Expanded AGN Sample Properties cont.}  \\ 
 \hline\hline
\multicolumn{7}{l}{\bf {\it 2d:  Radio Quiet Quasars}}  \\ 
 \hline

\multicolumn{1}{c}{Galaxy Name} & \multicolumn{1}{c}{{\it z}} & 
log(M$_{\rm BH}$)$^a$ & log(L$_{\rm bol}$)$^a$ & log(L$_{\rm FIR}$) & log( L$_{\rm FIR}$/L$_{\rm B}$) & log(L$_{\rm bol}$/L$_{\rm Edd}$)\\

\multicolumn{1}{c}{(1)} & \multicolumn{1}{c}{(2)} & (3) & (4) & (5) & (6) & 
(7)\\ 
\hline\hline
PG 0157+001 & 0.164 & 7.70 & 45.6 & 12.1$^{\dagger}$ & 1.17 & -0.18\\
PG 0923+201 & 0.190 & 8.94 & 46.2 & 11.7 & 1.02 & -0.82\\
PG 1202+281 & 0.165 & 8.29 & 45.4 & 11.1 & 0.42 & -1.00\\
PG 1402+261 & 0.164 & 7.29 & 45.1 & 11.1$^{\dagger}$ & 0.40 & -0.26\\
PG 1444+407 & 0.267 & 8.06 & 45.9 & 11.2$^{\dagger}$ & 0.26 & -0.23\\
PG 0804+761 & 0.100 & 8.24 & 45.9 & 10.6$^{\dagger}$ & 0.15 & -0.41\\
PG 1211+143 & 0.085 & 7.49 & 45.8 & 11.0 & 0.46 & 0.22\\
PG 1229+204 & 0.064 & 8.56 & 45.0 & 10.4 & 0.35 & -1.65\\
PG 1351+640 & 0.087 & 8.48 & 45.5 & 11.2 & 0.64 & -1.08\\
PG 1411+442 & 0.089 & 7.57 & 45.6 & 10.4 & 0.04 & -0.09\\
PG 1426+015 & 0.086 & 7.92 & 45.2 & 10.8 & 0.39 & -0.83\\
PG 1613+658 & 0.129 & 8.62 & 45.7 & 11.5$^{\dagger}$ & 1.02 & -1.06\\
PG 1617+175 & 0.114 & 7.88 & 45.5 & 10.8 & 0.40 & -0.46\\
PG 1700+518 & 0.292 & 8.31 & 46.6 & 11.9$^{\dagger}$ & 0.63 & 0.15\\
PG 2130+099 & 0.061 & 7.74 & 45.5 & 10.6 & 0.37 & -0.37\\

\hline\hline
\end{tabular}
\end{center}
\end{table*}

\begin{table*}
\begin{center}
\begin{tabular}{lcccccc}
\multicolumn{7}{l}{\bf Table 2: Expanded AGN Sample Properties cont.}  \\ 
 \hline\hline
\multicolumn{7}{l}{\bf {\it 2e:  Radio Loud AGN}}  \\ 
 \hline

\multicolumn{1}{c}{Galaxy Name} & \multicolumn{1}{c}{{\it z}} & 
log(M$_{\rm BH}$)$^b$ & log(L$_{\rm bol}$)$^b$ & log(L$_{\rm FIR}$) & log( L$_{\rm FIR}$/L$_{\rm B}$) & log(L$_{\rm bol}$/L$_{\rm Edd}$)\\

\multicolumn{1}{c}{(1)} & \multicolumn{1}{c}{(2)} & (3) & (4) & (5) & (6) & 
(7)\\ 
\hline\hline
3C 031 & 0.017 & 7.89 & 42.0 & 9.8 & 0.09 & -4.00\\
3C 84 & 0.018 & 9.28 & 44.0 & 10.7 & 0.62 & -3.35\\
3C 338 & 0.030 & 9.23 & 42.3 & 9.8 & -0.60 & -4.99\\
3C 465 & 0.030 & 9.32 & 42.6 & 9.8 & -0.34 & -4.81\\
3C 088 & 0.030 & 8.53 & 42.4 & 9.9 & 0.34 & -4.20\\
3C 285 & 0.079 & 8.46 & 41.5 & 10.8 & 0.80 & -5.08\\
3C 327 & 0.105 & 9.00 & 42.4 & 11.2 & 0.93 & -4.72\\
3C 390.3 & 0.056 & 8.41 & 44.9 & 10.3 & -0.01 & -1.64\\
3C 402 & 0.025 & 8.18 & 42.4 & 9.9 & 0.52 & -3.85\\
3C 47 & 0.425 & 8.52 & 45.6 & 12.0$^{\dagger}$ & 1.41 & -1.06\\
3C 138 & 0.759 & 8.73 & 45.9 & 12.3$^{\dagger}$ & 1.78 & -0.96\\
3C 249.1 & 0.312 & 9.48 & 45.7 & 11.1$^{\dagger}$ & -0.15 & -1.90\\
3C 298 & 1.436 & 10.15 & 47.4 & 13.2$^{\dagger}$ & 1.22 & -0.89\\
3C 380 & 0.692 & 9.37 & 46.4 & 11.9$^{\dagger}$ & 0.49 & -1.10\\

\hline\hline
\end{tabular}
\end{center}
\end{table*}

\begin{table*}
\begin{center}
\begin{tabular}{lcccccc}
\multicolumn{7}{l}{\bf Table 2: Expanded AGN Sample Properties cont.}  \\ 
 \hline\hline
\multicolumn{7}{l}{\bf {\it 2f:  Narrow Line Type 1 Seyferts}}  \\ 
 \hline

\multicolumn{1}{c}{Galaxy Name} & \multicolumn{1}{c}{{\it z}} & 
log(M$_{\rm BH}$)$^{\star}$ & log(L$_{\rm bol}$)$^c$ & log(L$_{\rm FIR}$) & log( L$_{\rm FIR}$/L$_{\rm B}$) & log(L$_{\rm bol}$/L$_{\rm Edd}$)\\

\multicolumn{1}{c}{(1)} & \multicolumn{1}{c}{(2)} & (3) & (4) & (5) & (6) & 
(7)\\ 
\hline\hline
NGC 4051 & 0.002 & 5.13 & 35.6 & 9.1 & 1.18 & -0.68\\
Mrk 142 & 0.045 & 6.77 & 37.6 & 10.2 & 0.80 & -0.31\\
Mrk 335 & 0.026 & 6.90 & 38.2 & $<$9.8 & $\cdots$ & 0.21\\
Mrk 478 & 0.077 & 7.44 & 38.9 & 10.9 & 0.65 & 0.34\\
Mrk 493 & 0.032 & 6.31 & 37.6 & 10.3 & 0.62 & 0.22\\
Mrk 684 & 0.046 & 6.80 & 38.4 & 10.4 & 0.63 & 0.45\\
Mrk 766 & 0.013 & 6.29 & 37.2 & 10.2 & 1.07 & -0.16\\
Mrk 1044 & 0.017 & 6.34 & 37.2 & 9.6 & 0.61 & -0.24\\
Ton S 180 & 0.062 & 6.85 & 38.7 & $<$10.4 & $<$0.99 & 0.76\\
RX J0323.2-4931 & 0.071 & 6.85 & 37.6 & $<$10.5 & $<$0.95 & -0.33\\
RX J1034.6+3938 & 0.044 & 5.81 & 37.5 & 10.3 & $\cdots$ & 0.62\\
RX J2301.8-5508 & 0.140 & 7.44 & 38.5 & 11.3 & $\cdots$ & -0.07\\
PG 1244+026 & 0.049 & 6.08 & 37.8 & 10.2 & $\cdots$ & 0.57\\
MS 2254-36 & 0.039 & 6.60 & 37.3 & $<$10.3 & $\cdots$ & -0.43\\

\hline\hline
\end{tabular}
\end{center}

{\noindent {\scriptsize {\bf Columns Explanation:} Col(1):Common Source Names; Col(2): Redshift; Col(3): Log of the mass of central black hole (M$_{\odot}$), $^{\star}$ Mass of the Central black hole was calculated using the suggested formulation of Grupe \& Mathur 2004: log(M$_{\rm BH}$) = 5.17 + log(R$_{\rm BLR}$) + 2[log FWHM(H$\beta$)-3], where R$_{\rm BLR}$ is the radius of the broad-line region (BLR) and is in units of light days. R$_{\rm BLR}$ may be calculated using the monochromatic luminosity at 5100 $\AA$ ($\lambda$L$_{5100}$) in units of watts:  log(R$_{\rm BLR}$) = 1.52 + 0.70(log($\lambda$L$_{5100}$)-37).  Monochromatic luminosities at 5100 $\AA$ were taken from Grupe \& Mathur 2004; Col(4): Log of the bolometric Luminosites in ergs s$^{-1}$; Col(5): Log of the far-infrared luminosities (in units of solar luminosities: L$\odot$) correspond to the 40-500$\mu$m wavelength interval and were calculated using the IRAS 60 and 100 $\mu$m fluxes according to the prescription: L$_{FIR}$=1.26$\times$10$^{-14}$(2.58f$_{60}$+f$_{100}$) in W m$^{-2}$ (Sanders \& Mirabel 1996), except $^{\dagger}$ where FIR luminosities correspond to rest-frame values taken from Haas et al. (2003) or Haas et al. (2004); Col(6): L$_{B}$: B magnitude taken from the Nasa/Ipac Extragalactic Database (NED); Col(7): Log of the Eddington ratio.}}
{\noindent{\scriptsize {\bf References: }$^{a}$ Woo \& Urry 2002 and references therein;  $^{b}$ Marchesini, Celotti, \& Ferrarese 2004 and references therein; $^{c}$ Grupe \& Mathur 2004. $^+$ Includes only those AGN-LINERs with black hole mass estimates from Papers I and II.  Combined with the AGN-LINERs in the archival LINER sample listed in Table 1, this represents the ``comprehensive AGN-LINER sample discussed in the text.}}
\end{table*}

\section{Observations and Data Reduction Procedure}

\subsection{\it Chandra Observations}
Archival {\it Chandra} observations of all LINER targets presented in this paper were obtained with the Advanced CCD Imaging Spectrometer (ACIS-S) with the source at the nominal aim point of the S3 CCD.  Most observations were carried out using the standard 3.2s frame time, although a few bright objects (NGC 3718 \& NGC 4258)  were observed with shorter frame times (0.44 s).  The exposure times vary for each object and are listed in Table 3.  In some cases, background flares were detected and had to be subtracted from the original exposure, resulting in shorter exposure times for these objects. For all observations the nuclear counts were insufficient for photon pileup to be significant. 

\begin{table*}
\begin{center}
\begin{tabular}{lccccc}
\multicolumn{6}{l}{\bf Table 3: {\it Chandra} Observation Log} \\ 
\hline\hline
\multicolumn{1}{c}{Galaxy} & OID & Exposure & R. A. & DEC. & Coordinate \\
\multicolumn{1}{c}{Name} & & Time & & & Catalog \\
\multicolumn{1}{c}{(1)} & (2) & (3) & (4) & (5) & (6)\\
\hline\hline
\multicolumn{6}{l}{{\it AGN-LINERs}}\\
\hline
NGC0315 & 4156 & 54127 & 00 57 48.887 & +30 21 08.84 & {\it 2MASS} \\
NGC2681 & 2060 & 79579 & 08 53 32.751 & +51 18 49.38 & {\it VLA} \\
NGC3169 & 1614 & 1953 & 10 14 15.36 & +03 27 57.40 & {\it VLA}\\
NGC3245 & 2926 & 9633 & 10 27 18.389 & +28 30 26.59 & {\it VLA} \\
NGC3718 & 3993 & 4911 & 11 32 34.848 & +53 04 4.56 & {\it VLA} \\
NGC4410A & 2982 & 34721 & 12 26 28.20 & +09 01 10.80 & {\it 2MASS} \\
NGC4258 & 2340 & 693512 & 18 57.533 & +47 18 14.06 & {\it VLA} \\
NGC4261 & 834 & 31465 & 12 19 23.227 & +05 49 29.89 & {\it VLA} \\
NGC4457 & 3150 & 36433 & 12 28 59.022 & +03 34 14.58 & {\it VLA}\\
NGC4552 & 2072 & 53492 & 12 35 39.804 & +12 33 22.91 & {\it VLA} \\
NGC4565 & 3950 & 54495 & 12 36 20.772 & +25 59 15.78 & {\it VLA} \\
NGC6482 & 3218 & 18430 & 17 51 48.81 & +23 04 19.0 & {\it 2MASS} \\
3C218 & 576 & 18364 & 09 18 05.675 & -12 05 44.30 & {\it VLA}\\
 \hline\hline
\multicolumn{6}{l}{\it NONAGN-LINERs}\\
\hline
NGC2541 & 1635 & 1927 & 08 14 40.07 & +49 03 41.2 & {\it 2MASS} \\
NGC2683 & 1636 & 1738 & 08 52 41.292 & +33 25 18.74 & {\it VLA} \\
NGC4150 & 1638 & 1738 & 12 10 33.3 & +30 24 05.50 & {\it VLA} \\
NGC4410B & 2982 & 34721 & 12 26 29.59 & +09 01 09.4 & {\it 2MASS} \\
NGC4438 & 2883 & 25073 & 12 27 45.567 & +13 00 32.87 & {\it VLA} \\
NGC4459 & 2927 & 9835 & 12 28 59.976 & +13 58 43.47 & {\it VLA} \\
NGC4501 & 2922 & 13823 & 12 31 59.175 & +14 25 12.98 & {\it VLA} \\
NGC4548 & 1620 & 2655 & 12 35 26.43 & +14 29 46.8 & {\it 2MASS} \\
NGC4550 & 1621 & 1880 & 12 35 30.60 & +12 13 15.3 & {\it 2MASS} \\
NGC4736 & 808 & 47366 & 12 50 53.064 & +41 07 13.65 & {\it VLA} \\
NGC5846 & 788 & 24091 & 15 06 29.294 & +01 36 20.39 & {\it VLA} \\
NGC5866 & 2879 & 23686 & 15 06 29.475 & +55 45 47.60 & {\it VLA} \\
\hline
\end{tabular}
\end{center}
{\scriptsize {\bf Column Explanation:} Col(1): Galaxy Common Name; Col(2): {\it Chandra} Observation Identification Number; Col(3): Exposure time in seconds; Col(4):  Right Ascension of nucleus in hours, minutes, \& seconds taken from the source in Column 6;  Col(5): Declination of nucleus in degrees, minutes, \& seconds, taken from the source in Column 6;  Col(6): {\it VLA} Coordinates or {\it 2MASS} coordinates used when extracting counts. {\it VLA} Coordinates come from the First Cataloge search, http://sundog.stsci.edu/cgi-bin/searchfirst.  {\it 2MASS} Coordinates came from NED.}
\end{table*}

The {\it Chandra} data were processed using CIAO v.3.0 using the latest calibration files provided by the {\it Chandra} X-ray Center (CXC).  In Table 4 we list the details of the {\it Chandra} observations. Our data reduction procedure and analysis follows closely the treatment described in our previous work (Paper I and Paper II).  The 0.3-8 keV energy range was chosen for analysis.  Nuclear count rates were extracted from a circular region of radius 2" centered on the nucleus, where the position was determined by radio observations, if available, or near-IR observations from the Two Micron All Sky Survey (2MASS).  The coordinates used for the extraction are listed in Table 3.  The background was extracted from a nearby circular region of radius 30" free of spurious X-ray sources.  The count rates and relative uncertainties were calculated using the procedure adopted in Paper I and Paper II.  Detections are defined when the X-ray counts are at least 3$\sigma$.  The upper limits listed in Table 4 are 3$\sigma$ values corresponding to a 2" extraction centered on the radio or infrared nucleus, with the background estimated as discussed above for the case of the detections.  Consistent with our definitions in Paper II, in this paper we define all AGN-LINERs as those galaxies with hard nuclear point sources coincident with the {\it VLA} or {\it 2MASS} nucleus and a 2-10keV luminosity $\geq$ 2$\times$10$^{38}$ ergs s$^{-1}$.

For fourteen of the twenty-five sources the detected counts in the 0.3-8 keV range were sufficient to perform detailed spectral fits.  Of these fourteen objects, eleven are classified as AGN-LINERs.  Spectral fitting was performed using XSPEC v.11.2.0 (Arnaud 1996).  We began by fitting all spectra with a single power-law model with the absorption column density fixed at the Galactic value.  This base model was then modified with additional components until an acceptable {\it $\chi$}$^{2}$$_{\rm red}$ was obtained.   We discuss these models in Section 4.1.  In the case of objects for which detected counts in the 0.3-8 keV regime were insufficient for spectral fitting, the nuclear count rate was converted to 2-10 keV X-ray luminosities assuming a canonical intrinsic power-law spectrum with photon index $\Gamma$=1.8 using the Galactic interstellar absorption listed in Table 4\footnote[2]{We adopt the sample power-law slope as Ho et al. 2001 which is typical for low luminosity AGN ($\Gamma$ ranges from 1.6-2.0; Paper 2; Terashima \& Wilson 2003)}.  We note here that several of the galaxies in our archival sample have been previously analyzed and published by various authors (e.g. Worrall et al. 2003, Donato et al. 2004, Terishima \& Wilson, 2003). We have reanalyzed all of these galaxies in a homogeneous way in order to ensure consistancy in the data reduction proceedure.

\subsection{\it ISOPHOT-S Observations}

We  searched  the {\it  ISOPHOT-S}  archive  for  6.2 $\mu$m  emission feature observations  of all galaxies presented in  this work.  Twenty one objects with previously unpublished 6.2 $\mu$m feature fluxes were found.  PHT-S consists of a dual grating spectrometer with a resolving
power of  90 (Laureijs et al. 2003).  Band SS covers the  range 2.5 -
4.8 $\mu$m,  while band SL  covers the range  5.8 - 11.6  $\mu$m.  The spectra for all but two  galaxies were obtained by operating the PHT-S aperture  in rectangular  chopping mode  - NGC  4102 and  Mrk  79 were observed with PHT-S in  triangular chopping mode.  For the rectangular chopping mode, the satellite pointed  to a position between the source and an  off-source position, and the chopper  moved alternatively between  these two  positions. The  triangular chopping  mode,  on the other   hand,   pointed  the   $24''\times24''$   aperture  of   PHT-S alternatively towards  the peak of  the emission and then  towards two background  positions off  the galaxy.  The source  was always  in the positive beam  in the spacecraft Y-direction.  The  calibration of the spectra was performed using  a spectral response function derived from several calibration stars of  different brightness observed in chopper
mode  (Acosta-Pulido   et  al.  2000).    The  relative  spectrometric
uncertainty  of the PHT-S  spectrum is  10\% when  comparing different parts of  the spectrum that  are more than  a few $\mu$m  apart.  The absolute  photometric  uncertainty  is  10\%  for  bright  calibration
sources.   All PHT  data processing  was performed  using  the ISOPHOT
Interactive Analysis  (PIA) system, version 10.0  (Gabriel 2002). Data reduction consisted  primarily of the removal  of instrumental effects such as radiation events which result in a increase of two consecutive read-out voltage  values.  The disturbance  is usually very  short and the slope of the ramp after  the glitch is similar to the slope before
it.   Once  the  instrumental  effects had  been  removed,  background
subtraction  was  performed and 6.2 $\mu$m line fluxes  were determined by integrating the  flux above the best  fit   linear  continuum  in   the  5.86-6.54$\mu$m  (rest-frame) wavelength interval.

The 24$\arcsec  \times 24\arcsec$ ISOPHOT aperture  corresponds to the central 0.5 kpc  for the nearest source to $\leq$ 60  kpc for the most
distant sources.   In 87\% of  the sources with published  or archival
6.2 $\mu$m  PHT-S observations,  the central 1  kpc nuclear  region is contained within the ISOPHOT beam.  Nuclear star formation in the vast majority of  galaxies originates from the central  few hundred parsecs (e.g. Surace \&  Sanders 1999; Scoville et al.  2000).  Indeed the MIR continuum  and PAH  emission as  well as  the FIR  flux is  within the ISOPHOT beam  for a  large number of  galaxies of  comparable redshift range to our  sample (e.g Lutz, Veilleux \&  Genzel, 1999, Lutz et al.,1998,  Rigopoulou  et al.,  1999).   Even  for  the nearest  galaxies,
imaging observations by Clavel et  al. (2000) of 14 spatially extended nearby  Seyferts  demonstrate  that,  on  average, 75\%  of  the  total 6.75$\mu$m continuum flux is contained within the ISOPHOT aperture.

\section{Results}

\subsection{ AGN Detection Rate and X-ray luminosities of AGN-LINERs}
In Paper II we defined AGN-LINERs as those objects which display a hard nuclear point source, with a 2-10 keV luminosity $\geq$ 2 $\times$ 10$^{38}$ ergs s$^{-1}$, coincident with the {\it VLA} or {\it 2MASS} nucleus. Of the 25 galaxies in the archival LINER sample, 13 meet this criterion.  Combining these statistics with those from Papers I and II, we find that 50\% (41/82) of LINERs have AGN nuclei.  As in Papers I and II, we assume that young supernova remnants, X-ray binaries, or hot diffuse gas from starburst driven winds are unlikely to be the source of the detections we observe, since these sources of emission are usually weak and/or spatially extended.  Moreover, detection of a single, dominant hard X-ray point source coincident with the {\it VLA} or {\it 2MASS} nucleus is highly suggestive of an AGN.

We assume in this work that LINERs that do not display dominant hard nuclear X-ray point sources do not harbor AGN (``{\it Non-AGN-LINERs}" listed in Tables 1, 3 \& 4).  These galaxies either display off-nuclear X-ray sources of comparable brightness to the nuclear source or no nuclear X-ray source.  Galaxies that display multiple sources emit primarily in the soft X-rays and are most likely to be morphologically consistent with starburst galaxies, expected to contain a large population of young stars and/or X-ray binaries.  Galaxies that lack a hard nuclear source either 1) lack an energetically significant AGN or 2) contain highly obscured AGN with column densities reaching $\sim$ 10$^{24}$ cm$^{-2}$.

From Table 4, the 2-10 keV luminosities of the AGN-LINERs in our archival LINER sample range from $\sim$ 2 $\times$ 10$^{38}$ to $\sim$ 1 $\times$ 10$^{42}$ ergs s$^{-1}$, well within the span of luminosities of the comprehensive AGN-LINER sample ($\sim$ 2 $\times$ 10$^{38}$ to $\sim$ 2 $\times$ 10$^{44}$ ergs s$^{-1}$). As in Paper II, the majority of objects occupy the 10$^{39}$ to 10$^{41}$ ergs s$^{-1}$ luminosity range.

The majority of the 2-10 keV X-ray luminosities listed in Table 4 were calculated assuming a generic power-law model with photon index $\Gamma$ = 1.8 using the Galactic absorption given in Table 4.  Fourteen of the twenty-five LINERs had counts sufficient for an XPEC spectral analysis, allowing us to gauge the accuracy of our generic power-law model.  Combing these fits with the 3 spectral fits from Paper II, we find that the photon index for AGN-LINERs ranged from $\sim$ 0.7 to 2.0 with an average of 1.6.  This range in photon index is consistent with the values found in a larger sample of low luminosity AGN (Terashima et al. 2002).  

Two of the fourteen galaxies (NGC 4410A \& 3C218 - both AGN-LINERs) were well fit by a single power-law model with absorption fixed at the Galactic value.   For seven of the fourteen galaxies (2 Non-AGN-LINERs and 5 AGN-LINERs) one additional component (either an intrinsic absorption or a thermal component) was required before an acceptable $\chi$ $_{\it red}^{2}$ was obtained. With respect to the thermal component, for all galaxies with low signal-to-noise (all except NGC 2681) the abundance was fixed at the Galactic value.  In the case of NGC 2681 the abundance was first left free to vary, which resulted in a best fit value of 0.8.  However, the spectral data for this galaxy were not good enough to estimate the errors on the abundance parameter when it was left free to vary.  The abundance for NGC 2681 was therefore fixed at the best fit value.  In these nine cases, the XSPEC luminosity was less than a factor of 4 different from the generic luminosities calculated using our generic power-law model.  The X-ray spectra of the AGN-LINERs NGC 4258, NGC 0315, \& NGC 4261, on the other hand, were well fit with more complex models that are described and shown below.   In the case of NGC 0315 and NGC 4261, the XSPEC luminosity differed by factors of only 3 and 4 respectively from the generic calculation.  However, in the case of NGC 4258, whose spectrum shows severe absorption in the soft band, the XSPEC-derived luminosity differs by a factor of nearly 15 from that derived using the generic power law model.   In addition, though the luminosities derived through the XSPEC and generic models roughly agree in the case of NGC 4261, the flat power law component, $\Gamma$=0.71, differs greatly from the $\Gamma$=1.8 described by our generic model.  These results lead us to conclude that our generic model is appropriate only in so far as the spectrum is simple, requiring few additional components.  Lastly, NGC 5846 (a Non-AGN-LINER) and NGC 6482 (an AGN-LINER) were fit using a thermal component and absorption fixed at the Galactic value.  In both cases, the spectra in the hard band (2-10 keV) were poorly constrained and characterized by large errors.   The fits for NGC 5846 and NGC 6482 are representative of the 0.3-2.0 keV band {\it only} and say nothing about the hard band (2-10 keV) luminosity associated with the power-law component and thus the accuracy of the luminosity derived from the generic power-law model.  We therefore choose to adopt the luminosities derived from the generic power-law model for these two galaxies until better spectral data for the two galaxies can be obtained.  The specific parameters and models for these fourteen galaxies are given in Table 5.

\subsection{ Spectral Fits for Individual Objects}

{\it NGC 4258}:  This galaxy's spectrum was initially fit with a single power-law model with $\Gamma$ = 1.4$^{+0.5}_{-0.3}$ and absorption column density fixed at the Galactic value. The resulting poor fit ({\it $\chi$}$^{2}$$_{\rm red}$ = 2.1, 70 d.o.f.) in addition to the clear absence of soft counts indicated the need for an additional absorption component. The resulting intrinsic absorption for the best fit model was {\it N}$_{\rm H}$ = (6.6 $^{+2.1}_{-1.3}$) $\times$ 10$^{22}$ cm$^{-2}$.  The model was further improved with the addition of a thermal component (kT = 0.83$^{+0.4}_{-0.7}$ keV, {\it Z}/{\it Z}$_{\odot}$ = 1.0), which is significant at greater than the 99\% confidence level.  This model yielded an acceptable fit ({\it $\chi$}$^{2}$$_{\rm red}$ = 0.50 (67 d.o.f.)).  The spectrum for this galaxy is shown in Figure 2.

{\it NGC 4261}: The spectrum for this galaxy was initially fit with a single power law model with $\Gamma$ = 0.7$^{+0.8}_{-0.7}$ and absorption column density fixed at the Galactic value. The resulting poor fit ({\it $\chi$}$^{2}$$_{\rm red}$ = 7.1, 71 d.o.f.) in addition to the clear absence of soft energies, as well as a large deficit in the 1.4 to 4 keV energy range indicated the need for two additional components. As per Gliozzi et al. 2003, the model was improved with the addition of a thermal component (kT = 0.61$^{+0.03}_{-0.02}$ keV, {\it Z}/{\it Z}$_{\odot}$ =1.0) as well as a partial covering fraction ({\it N}$_{\rm H}$ = (5.3 $^{+3.7}_{-2.8}$) $\times$ 10$^{22}$ cm$^{-2}$, Covering Fraction = 85$^{+0.2}_{-0.6}$\%), both of which are significant at the 99\% confidence level.  This final model yielded an acceptable fit ({\it $\chi$}$^{2}$$_{\rm red}$ = 0.69 (67 d.o.f.)).  The spectrum for this galaxy is shown in Figure 3.

{\it NGC 0315}: This galaxy's spectrum was initially fit with a single power law model with $\Gamma$ = 1.6$^{+0.1}_{-0.2}$ and absorption column density fixed at the Galactic value. The resulting poor fit ({\it $\chi$}$^{2}$$_{\rm red}$ = 1.5, 166 d.o.f.) in addition to the clear absence of soft energies indicated the need for an additional absorption component. The resulting  intrinsic absorption for the best fit model was {\it N}$_{\rm H}$ = (8.0 $^{+1.0}_{-2.0}$) $\times$ 10$^{21}$ cm$^{-2}$.  The model was further improved with the addition of a thermal component (kT = 0.54$^{+0.3}_{-0.5}$ keV, {\it Z}/{\it Z}$_{\odot}$ = 1.0), which is significant at greater than 99\% confidence level.  This model yielded an acceptable fit ({\it $\chi$}$^{2}$$_{\rm red}$ = 0.58, 162 d.o.f.).  The spectrum and model for this galaxy is shown in Figure 4. 

Our spectral analysis suggests that for the vast majority of AGN-LINERs in the comprehensive AGN-LINER sample, our generic power-law model is likely to yield reasonable 2-10 keV luminosities. In those targets with accurate spectral fits (excluding NGC 5846 and NGC 6482), we have adopted the XSPEC-derived luminosities in all of our calculations and plots.

\begin{table*}
\begin{center}
\begin{tabular}{lccccccc}
\multicolumn{8}{l}{{\bf Table 4: Results of the Archival LINER Sample}} \\
\hline\hline
\multicolumn{1}{c}{Galaxy} & \multicolumn{1}{c}{{\it N}$_{\rm H}$} & Hard & Count & Hardness & log & log & log  \\
\multicolumn{1}{c}{Name} & \multicolumn{1}{c}{cm$^{-2}$} & Counts & Rate & Ratio &(L$_{\rm X-Generic}$) & (L$_{\rm X-Xspec}$) & L$_{\rm bol}$/L$_{\rm Edd}$ \\
\multicolumn{1}{c}{(1)} & \multicolumn{1}{c}{(2)} & (3) & (4) & (5) & (6) & 
(7) & (8)\\ 
\hline\hline
\multicolumn{8}{l}{\it AGN-LINERs}\\
\hline
NGC0315 & 5.87 & 1862 & 0.084$\pm$0.001 & -0.18 & 41.22 & 41.67 & -4.02 \\
NGC2681 & 2.42 & 103 & 0.012$\pm$0.001 & -0.78 & 38.92 & 38.74 & -5.89 \\
NGC3169 & 2.66 & 150 & 0.080$\pm$0.007 & 0.92 & 40.09 & $\cdots$ & -4.49 \\
NGC3245 & 2.13 & 17 & 0.007$\pm$0.001 & -0.53 & 36.16 & $\cdots$ & -5.98 \\
NGC3718 & 1.08 & 609 & 0.227$\pm$0.007 & 0.09 & 40.44 & 40.06 & -3.65 \\
NGC4258 & 1.51 & 1541 & 0.239$\pm$0.006 & 0.86 & 39.62 & 40.80 & -3.04 \\
NGC4261 & 1.52 & 545 & 0.069$\pm$0.002 & -0.50 & 40.51 & 41.15 & -4.41 \\
NGC4410A & 1.71 & 251 & 0.036$\pm$0.001 & -0.60 & 41.14 & 41.24 & -4.06 \\
NGC4457 & 1.80 & 61 & 0.013$\pm$0.001 & -0.74 & 39.22 & 39.22 & -4.64 \\
NGC4552 & 2.57 & 219 & 0.032$\pm$0.001 & -0.74 & 39.55 & 39.62 & -6.70 \\
NGC4565 & 1.32 & 543 & 0.035$\pm$0.001 & -0.43 & 39.09 & 39.34 & -4.29 \\
NGC6482$^a$ & 7.67 & 17 & 0.022$\pm$0.001 & -0.92 & 40.47 & $\cdots$ & -5.02 \\
3C218 & 4.93 & 202 & 0.033$\pm$0.001 & -0.34 & 41.82 & 42.15 & -3.30 \\

\hline\hline
\multicolumn{8}{l}{\it NONAGN-LINERs} \\
\hline

NGC2541 & 4.60 & $<$1 & $<$0.001 & $\cdots$ & $<$37.60 & $\cdots$ & $\cdots$ \\
NGC2683 & 3.10 & 14 & 0.009$\pm$0.002 & 0.87 & 38.06 & $\cdots$ & $\cdots$ \\
NGC4150 & 1.63 & $<$1 & $<$0.001 & $\cdots$ & $<$37.16 & $\cdots$ & $\cdots$ \\
NGC4410B & 1.71 & 1 & $<$0.001 & -0.80 & 38.51 & $\cdots$ & $\cdots$ \\
NGC4438 & 2.64 & 53 & 0.028$\pm$0.001 & -0.85 & 39.50 & 37.54 & $\cdots$ \\
NGC4459 & 2.72 & 9 & 0.006$\pm$0.001 & -0.70 & 38.84 & $\cdots$ & $\cdots$ \\
NGC4501 & 2.47 & 30 & 0.006$\pm$0.001 & -0.33 & 38.90 & $\cdots$ & $\cdots$ \\
NGC4548 & 2.35 & 17 & 0.010$\pm$0.002 & 0.26 & 39.05 & $\cdots$ & $\cdots$ \\
NGC4550 & 2.60 & $<$2 & $<$0.001 & $\cdots$ & $<$37.68 & $\cdots$ & $\cdots$ \\
NGC4736 & 1.43 & 446 & 0.068$\pm$0.001 & -0.72 & 38.68 & 38.72 & $\cdots$ \\
NGC5846$^a$ & 4.27 & 7 & 0.008$\pm$0.001 & -0.92 & 39.44 &$\cdots$ & $\cdots$ \\NGC5866 & 1.47 & 22 & $<$0.001 & 0.34 & 37.69 & $\cdots$ & $\cdots$ \\

\hline\hline
\end{tabular}
\end{center}
{\scriptsize{\bf Notes:} {\it a} Luminosities derived from the generic power-law model rather than the Xspec model are adopted for these two galaxies in all of our plots and calculations.  See Section 4.1 for details.}
{\scriptsize{\bf Columns Explanation:} Col(1):Common Source Names; Col(2): Galactic {\it N}$_{\rm H}$(in units of$\times $ 10$^{20}$ cm$^{-2}$); Col(3): Hard counts in the nucleus (2-8 keV) from an extraction region of radius 2'' centered on the radio or {\it 2MASS} nucleus; Col(4): Countrate (counts/sec) in the 0.3-10 keV band; Col(5): Hardness Ratio: Defined here as (H-S)/(H+S), where H represents the hard counts (2-8 keV) and S represents the soft counts (0.3-2 keV) in the nucleus; Col(6):  Log of the X-ray luminosity (2-10keV) in ergs s$^{-1}$ calculated using our generic model; Col(7): Log of the X-ray luminosity (2-10keV) in ergs s$^{-1}$ calculated using the Xspec model; Col(8): Log of the Eddington ratio.}
\end{table*}

\begin{table*}
\begin{center}
\begin{tabular}{lccccc}
\multicolumn{6}{l}{\bf Table 5: Results of Spectral Analysis}  \\
\hline\hline
\multicolumn{1}{c}{Galaxy Name} & \multicolumn{1}{c}{Model$^b$} & kT & N$_{\it H}^{\it Add}$ & $\Gamma$ & $\chi$$_{\it red}^{2}$/d.o.f.\\

\multicolumn{1}{c}{(1)} & \multicolumn{1}{c}{(2)} & (3) & (4) & (5) & (6) 
 \\
\hline
NGC0315$^a$ & IV & 0.54$_{-0.05}^{+0.03}$ & 0.8$_{-0.2}^{+0.1}$ &  1.60$_{-0.24}^{+0.14}$ & 0.58/162\\

NGC2681$^a$ & III & 0.73$\pm$0.06 & $\cdots$ &  1.57$_{-0.57}^{+0.47}$ & 0.47/29\\

NGC3718$^a$ & II & $\cdots$ & 0.8$_{-0.3}^{+0.4}$ & 1.48$_{-0.35}^{+0.42}$ & 0.43/45\\

NGC4258$^a$ & IV & 0.83$_{-0.73}^{+0.37}$ & 6.6$_{-1.3}^{+2.1}$ & 1.44$_{-0.34}^{+0.52}$ & 0.50/67\\

NGC4261$^{a,c}$ & VI & 0.61$_{-0.02}^{+0.03}$ & 5.3$_{-2.8}^{+3.7}$ & 0.71$_{-0.71}^{+0.80}$ & 0.69/67\\

NGC4410A$^a$ & I & $\cdots$ & $\cdots$ & 1.73$\pm$0.14 & 0.36/51\\

NGC4438 & III & 0.77$\pm$0.06 & $\cdots$ & 1.19$_{-0.48}^{+0.67}$ & 0.93/22\\

NGC4457$^a$ & III & 0.69$\pm$0.13 & $\cdots$ & 1.57$_{-0.57}^{+0.70}$ & 0.38/15\\

NGC4552$^a$ & III & 0.75$_{-0.08}^{+0.06}$ & $\cdots$ & 1.62$_{-0.14}^{+0.16}$ & 0.64/59\\

NGC4565$^a$ & II & $\cdots$ & 0.2$\pm$0.06 & 1.92$_{-0.21}^{+0.23}$ & 0.66/71\\

NGC4736 & III & 0.67$_{-0.06}^{+0.07}$ & $\cdots$ & 1.60$_{-0.08}^{+0.09}$ & 0.48/46\\

NGC5846 & V & 0.65$_{-0.14}^{+0.17}$ &  $\cdots$ & $\cdots$ & 0.46/13\\

NGC6482$^a$ & V & 0.80$\pm$0.06 & $\cdots$ & $\cdots$ & 0.57/15\\

3C218$^a$ & I & $\cdots$ & $\cdots$ & 1.17$_{-0.23}^{+0.24}$ & 0.31/26\\

\hline
\end{tabular}
\end{center}
{\scriptsize {\bf Notes:} {\it a}: indicates AGN-LINERs,  {\it b}: I=wabs(powerlaw), II=wabs(zwabs(powerlaw)), III=wabs(apec+powerlaw), IV=wabs(apec+zwabs(powerlaw)), V=wabs(apec), VI=wabs(apec+zpcfabs(powerlaw)), {\it c} The fit for NGC 4261 also required a partial covering fraction of 0.85$_{-0.62}^{+0.15}$\%.}
{\scriptsize {\bf Columns Explanation:} Col.(1): Common Source Names, ;  Col.(2): Spectral model (see note); Col.(3): Plasma temperature in keV; Col.(4): Absorption column density at the redshift of the source in units of 10$^{22}$cm$^{-2}$; Col(5): Photon index; Col(6): reduced $\chi$$^2$ and degrees of freedom.}
\end{table*}

\begin{figure}[h]
\includegraphics[width=6cm,angle=270]{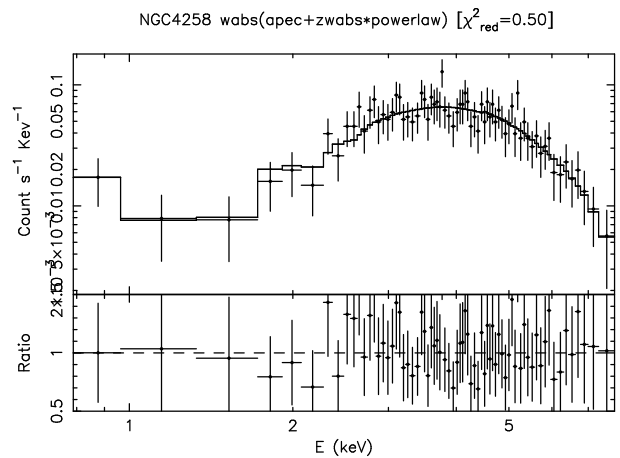}\\
\caption[]{Best fit model for Non-AGN LINER, NGC 4261}
\end{figure}

\begin{figure}[h]
\includegraphics[width=6cm,angle=270]{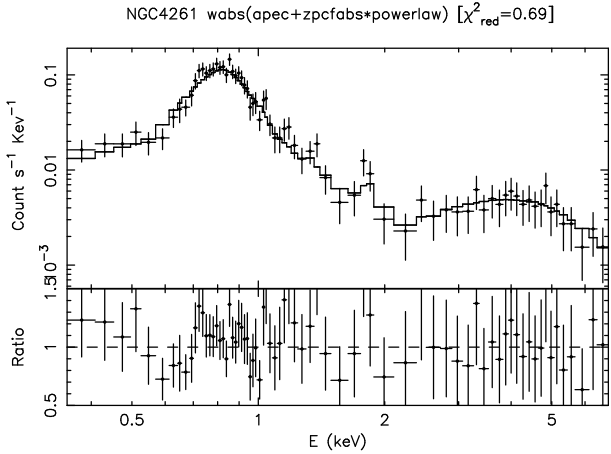}\\
\caption[]{Best fit model for AGN LINER, NGC 4258}
\end{figure}

\begin{figure}[h]
\includegraphics[width=6cm,angle=270]{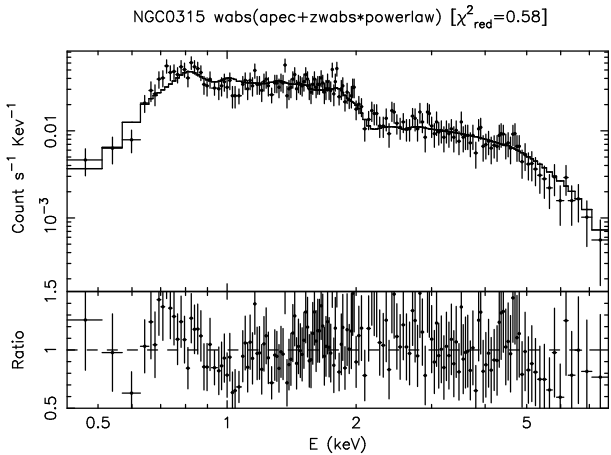}\\
\caption[]{Best fit model for AGN LINER, NGC 0315}
\end{figure}

\subsection{Correlation Between Eddington Ratio and FIR Luminosity and IR Brightness in AGN-LINERs}

In Figures 5a and 5b we plot L$_{\rm bol}$/L$_{\rm Edd}$  versus L$_{\rm FIR}$, L$_{\rm FIR}$/L$_{\rm B}$ for the 34 out of 41 AGN-LINERs in our comprehensive LINER sample that have black hole estimates.  As can be seen, the original correlations are reinforced by our expanded AGN-LINER sample, extending over seven orders of magnitude in L$_{\rm bol}$/L$_{\rm Edd}$.  Employing a Spearman rank correlation analysis (Kendall \& Stuart 1976) to assess the statistical significance of our correlation yields a  correlation coefficient of r$_{\rm S}$ = 0.50 between  L$_{\rm bol}$/L$_{\rm Edd}$ and L$_{\rm FIR}$/L$_{\rm B}$ with a probability of chance correlation of 2.9 $\times$ 10$^{-3}$, indicating a significant correlation.  In the case of L$_{\rm bol}$/L$_{\rm Edd}$ and L$_{\rm FIR}$, the Spearman rank test gives a correlation coefficient  of r$_{\rm S}$ = 0.62 with a probability of chance correlation of 7.8 $\times$ 10$^{-5}$, again indicating a significant correlation.  The Spearman rank correlation technique has the advantage of being non-parametric, robust to outliers and does not presuppose a linear relation. 

The correlations presented in Figures 5a and 5b appear to be primary and are unlikely to be induced by either distance effects or correlations between any individually observed quantities used to calculate the Eddington ratio.  The 2-10 keV X-ray and FIR fluxes for the entire sample of AGN-LINERs show no correlation (r$_{\rm S}$ = 0.03), suggesting that the correlations shown in Figures 5a and 5b are fundamental.  We also conducted a Spearman partial correlation analysis on our datasets to check whether spurious correlations may have been produced by distance effects. The partial correlation describes the relationship between two variables when the third variable is held constant. The partial Spearman rank correlation coefficient between L$_{\rm bol}$/L$_{\rm Edd}$  and L$_{\rm FIR}$, L$_{\rm FIR}$/L$_{\rm B}$ holding the distance fixed was higher in both cases (P$_{\rm S}$ = 0.69 and P$_{\rm S}$= 0.64, respectively), suggesting that the Eddington ratio and FIR luminosity and IR-brightness ratio in AGN-LINERs are indeed physically correlated quantities.

A formal fit to the plots in Figures 5a and 5b yields the following relationships:

\begin{equation}
\log(L_{\rm bol}/L_{\rm Edd}) =  (1.15 \pm 0.19)\log(L_{\rm FIR}) + (-14.92 \pm 1.75)
\end{equation}
\begin{equation}
\log(L_{\rm bol}/L_{\rm Edd}) = (1.13 \pm 0.22)\log(L_{\rm FIR}/L_{B}) + (-4.13 \pm 0.19)
\end{equation}

The dispersion in the plots displayed in Figures 5a and 5b is significant (rms scatter of 0.99 and 1.06 dex in L$_{\rm bol}$/L$_{\rm Edd}$ for Figures 5a and 5b, respectively).  In Paper II, we pointed out the difficulty in assessing how much of the scatter is intrinsic or due to the uncertainties in the derived quantities.  The uncertainty in the X-ray luminosity derived from our power-law model, the uncertainty in the bolometric correction factor for the LINER class, the uncertainty and uniform applicability of the M$_{\rm BH}$  vs. $\sigma$ relationship for this sample of LINERs, and the non-simultaneity of the observations will all introduce some scatter.  If we assume that the uncertainty in the black hole mass estimate is 0.3 dex (Tremaine et al. 2002) and that there is no X-ray, optical, or FIR variability in this sample of LINERs, then the uncertainty in L$_{\rm bol}$ would need to be a factor of $\sim$ 5 if the scatter is entirely due to the uncertainties in the derived quantities plotted in Figures 5a and 5b.  The bolometric luminosity in our sample of LINERs, estimated using the X-ray luminosity (see Section 2), is certainly uncertain within this factor, if not more.  The bolometric correction factor used in this work relies on the average SED of only 7 low luminosity AGN presented by Ho (1999) since more extensive studies of the SEDs of LINERs are nonexistent.  Indeed,  the correction factor for these 7 objects varies by a factor of 6 (L$_{\rm bol}$ = 11 $\times$ L$_{\rm X}$(2-10 keV) to L$_{\rm bol}$ = 69 $\times$ L$_{\rm X}$(2-10 keV)).  In addition, there is some uncertainty in adopting a generic power-law model to calculate the X-ray luminosity.  As described in section 4.1, detailed spectral fits of the X-ray spectrum of a few of the targets in the archival LINER sample show that the derived luminosities can deviate substantially from the generic power law-derived luminosities.  Given these uncertainties, it is entirely plausible that the scatter in Figures 5a and 5b is not intrinsic and can be completely attributable to the uncertainties in the derived quantities.

\begin{figure*}[]
{\includegraphics[width=8cm]{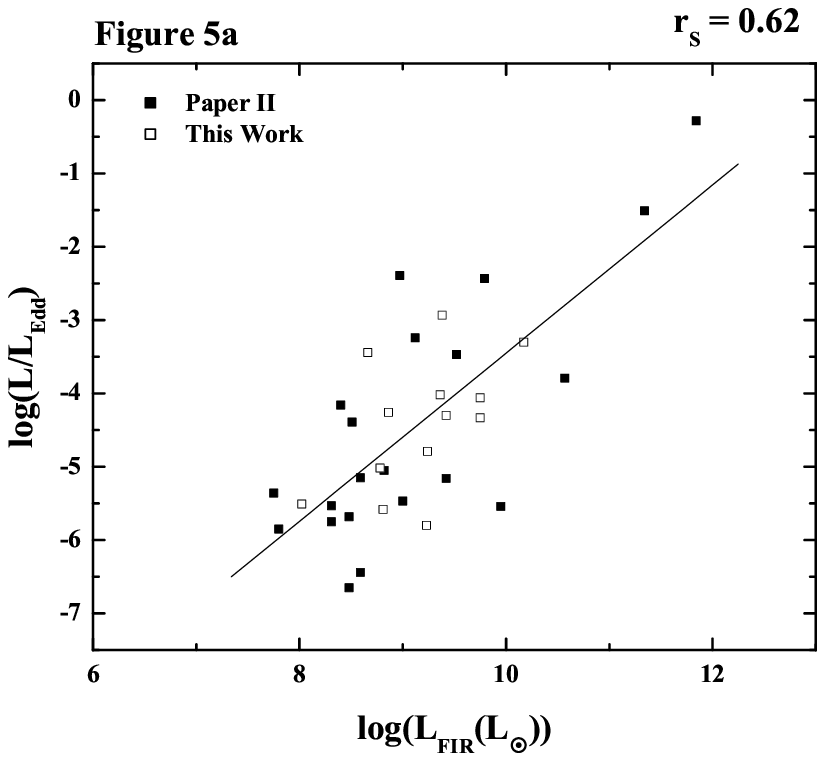}}{\includegraphics[width=9cm]{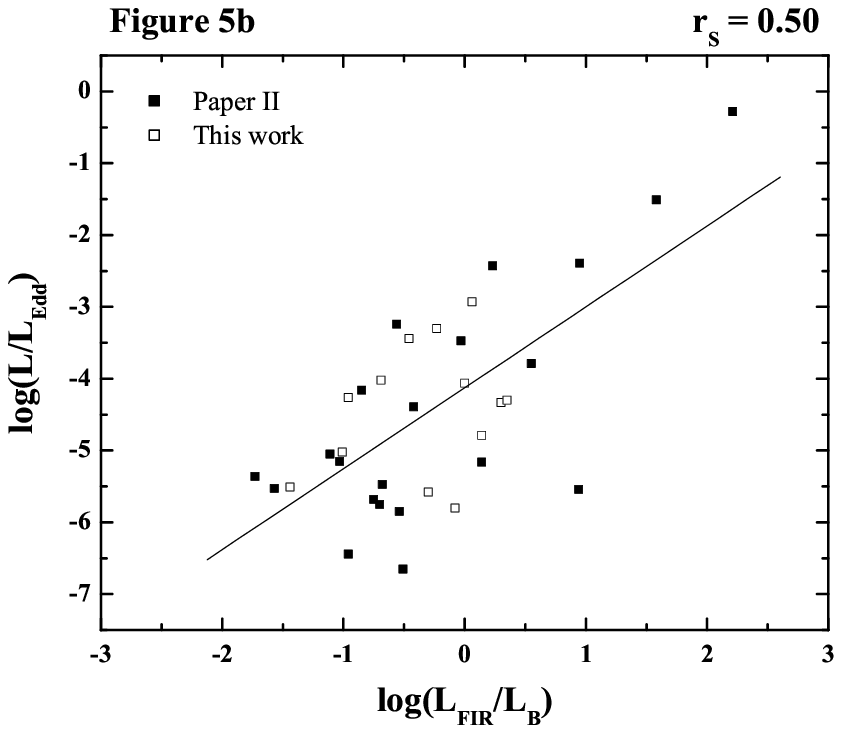}}\\
\caption[]{Correlation between L$_{\rm bol}$/L$_{\rm Edd}$  verses L$_{\rm FIR}$ (a) L$_{\rm FIR}$/L$_{\rm B}$ (b) for the entire set of 34/41 AGN-LINERs that have black hole estimates in our comprehensive LINER sample. There is a significant correlation in each plot that extends over seven orders of magnitude in L/L$_{\rm Edd}$.  The Spearman rank correlation coefficients are given in the upper right corner of each plot.}
\end{figure*}

\subsection{Correlation Between Eddington Ratio and FIR Luminosity in other AGN Subclasses}

In Figure 6 we plot L$_{\rm bol}$/L$_{\rm Edd}$  versus L$_{\rm FIR}$ for the entire expanded AGN sample included in this paper.  We highlight all targets with redshift z $>$ 0.1 as open symbols in the plot since distance effects may spuriously reinforce the correlation.  Interestingly, the original correlation is in general reinforced by our expanded AGN sample, extending in this case over almost nine orders of magnitude in L$_{\rm bol}$/L$_{\rm Edd}$.  For the various AGN subclasses, the origin of the optical luminosity is likely to vary tremendously.  For example, massive ellipticals are often the hosts of many radio-loud AGN and LINERs.  In these cases, a significant fraction of the blue luminosity is likely to originate from the host galaxy.  On the other hand, in RQQ for example, the bulk of the optical luminosity is likely to originate from an optically thick accretion disk.  We therefore choose to plot the FIR luminosity only in our investigation of possible correlations.  

Employing a Spearman rank correlation analysis to the entire 129 galaxies plotted in Figure 6 yields a  correlation coefficient of r$_{\rm S}$ = 0.65 between  L$_{\rm bol}$/L$_{\rm Edd}$ and L$_{\rm FIR}$ with a probability of chance correlation of 6.9 $\times$ 10$^{-17}$, indicating a significant correlation.  The best-fit linear relationship to the entire dataset is:

\begin{equation}
\log(L_{\rm bol}/L_{\rm Edd}) =  (1.22 \pm 0.14)\log(L_{\rm FIR}) + (-14.14 \pm 1.38)
\end{equation}

Apart from the overall correlation, there are correlations between the quantities plotted in Figure 6 for each AGN subclass.  We list the Spearman rank correlation coefficients between L$_{\rm bol}$/L$_{\rm Edd}$ and L$_{\rm FIR}$ for each AGN subclass in Table 6.    Figure 6 shows that the different AGN subclasses occupy distinct regions in the L$_{\rm bol}$/L$_{\rm Edd}$ and L$_{\rm FIR}$ plane.  Most notably, the Seyferts, RQQs, and NLS1s in general display a shallower slope and larger intercept than do the LINERs and radio-loud AGN.  For example, the best-fit linear relationship to the entire Seyfert dataset is:

\begin{equation}
\log(L_{\rm bol}/L_{\rm Edd}) =  (0.66 \pm 0.11)\log(L_{\rm FIR}) + (-7.81 \pm 1.09)
\end{equation}
which is significantly different from the best-fit linear relationship for the LINER dataset (Equation 1).  In addition, the scatter in the Seyfert dataset displayed in Figure 6 is significantly less than that in the LINER dataset (0.5 dex in L$_{\rm bol}$/L$_{\rm Edd}$ compared with 1.09 dex in the Seyfert and LINER datasets, respectively).  This may be the result of the fact that the bolometric luminosity, obtained by direct flux integration of a well-sampled SED, is much more reliable in the Seyferts compared with the LINERs.  

\begin{figure*}[]
{\includegraphics[width=16cm]{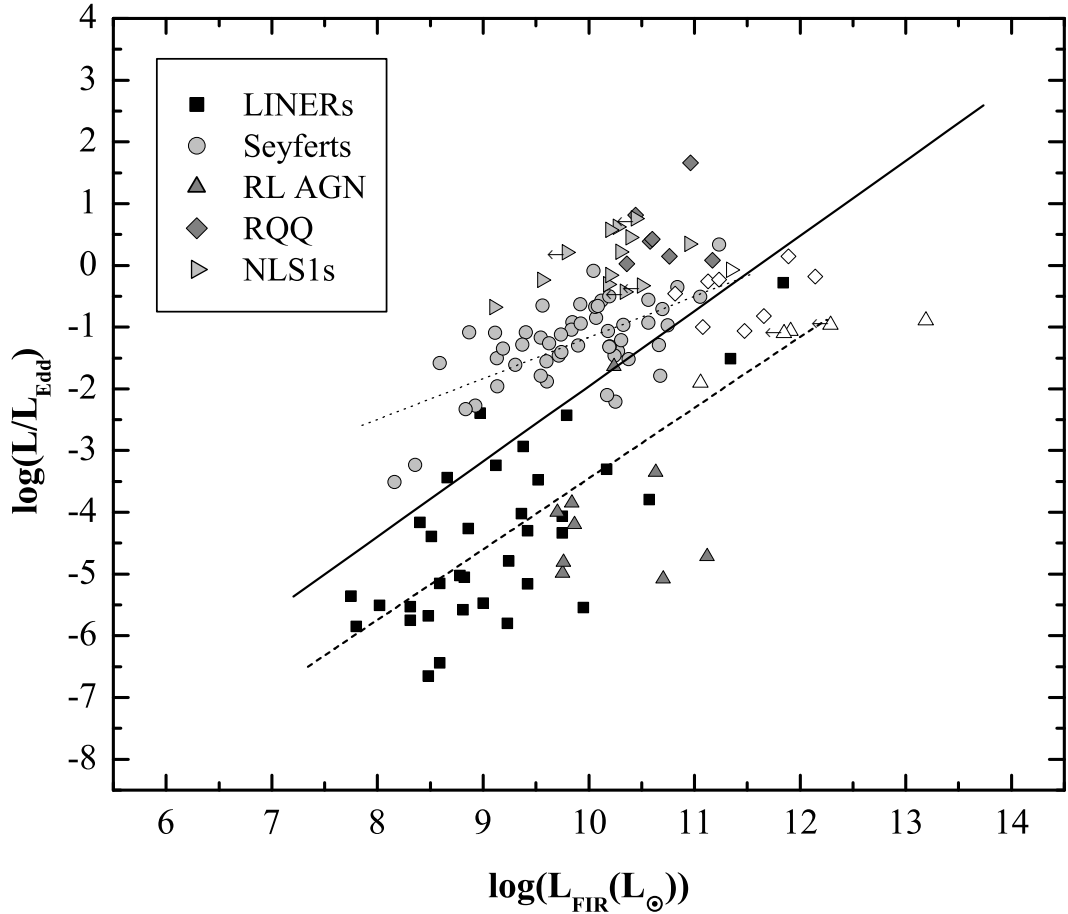}}\\
\caption[]{Correlation between L$_{\rm bol}$/L$_{\rm Edd}$  verses L$_{\rm FIR}$ for the entire set of AGN presented in this work. This includes the 34/41 AGN-LINERs that have black hole estimates. This plot shows a significant correlation that extends over almost nine orders of magnitude in L/L$_{\rm Edd}$.  The best-fit linear relationship to the entire dataset is displayed by the solid line.  The dashed and dotted lines correspond to the best-fit line applied only to the LINER, and Seyfert datasets, respectively.  The Spearman rank correlation coefficients for the different datasets are give in Table 6.}
\end{figure*}

\begin{figure}[]
{\includegraphics[width=9cm]{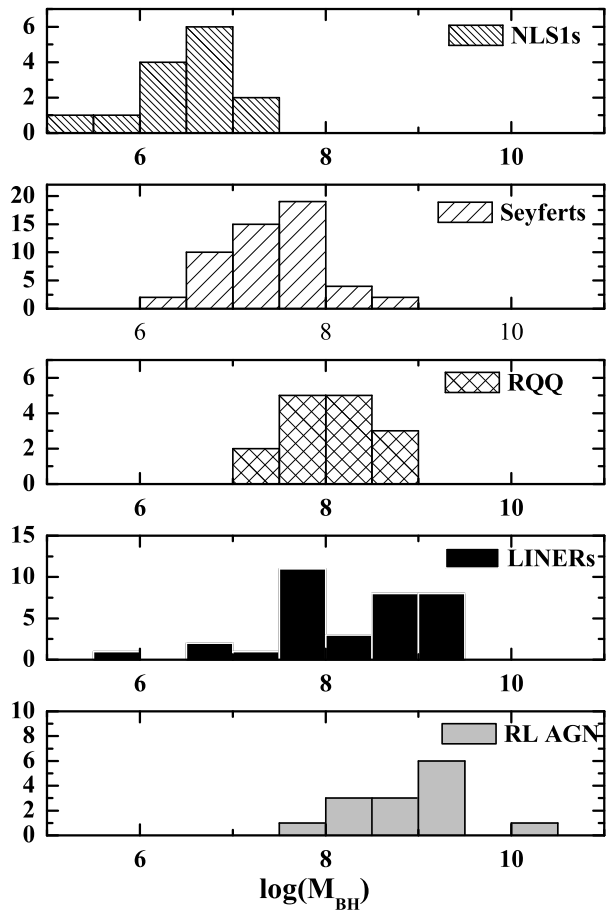}}\\
\caption[]{Black hole masses for the entire set of AGN presented in this work. }\end{figure}
\begin{figure}[]
{\includegraphics[width=9cm]{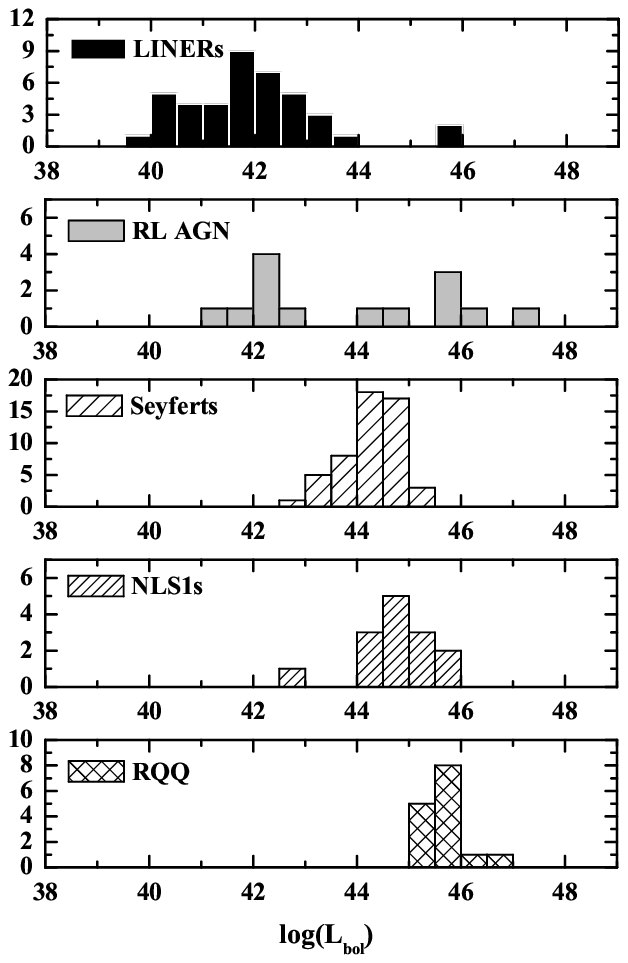}}\\
\caption[]{Bolemetric luminosities for the entire set of AGN presented in this work.}
\end{figure}

If the different AGN subclasses occupy distinct regions of the plot in Figure 6, this may be an important result.  This means that for a given L$_{\rm FIR}$, Seyferts, RQQ, and NLS1s tend to show higher values of L$_{\rm bol}$/L$_{\rm Edd}$ than LINERs.  A spurious result may be obtained if L$_{\rm bol}$ is systematically underestimated in the LINERs or if L$_{\rm Edd}$ i.e., M$_{\rm BH}$ - is systematically overestimated relative to the Seyferts, RQQ, and NLS1s.  Figures 7 \& 8 show the distributions of black hole masses and bolometric luminosities for each AGN subclass.  The plot clearly shows that each AGN subtype shows different distributions of black hole masses and bolometric luminosities.  Radio-loud AGN and LINERs tend to have more massive black holes than Seyferts and NLS1s have the least massive black holes.  A formal Kolmogorov-Smirnov (K-S) test between the black hole masses of Seyferts and LINERs, for example, demonstrates that the two distributions are significantly different, with a probability of being drawn from the same population of less than 0.00046\%.  This result is unlikely to be spurious.  NLS1s most likely contain lower mass black holes than standard Seyferts which is why they display narrow lines (e.g. Peterson et al. 2000; Wandel \& Boller (1998)).  Likewise, the hosts of radio-loud AGN and LINERs are often massive ellipticals, which are typically more massive than spirals - the hosts of most standard Seyferts (e.g. Ho, Filippenko, \& Sargent 2003).  Figure 9 shows that the bolometric luminosities of the LINERs are lower than that of the other AGN subclasses. A formal K-S test between the bolometric luminosities of Seyferts and LINERs, for example, demonstrates that the two distributions are significantly different, with a probability of being drawn from the same population of less than 3\%.  Again, this result is unlikely to be spurious.  Most of the LINERs in our sample are low luminosity AGN, which do contain less luminous nuclei than standard Seyferts (e.g. Koratkar et al. 1995, Terashima et al. 2003), radio galaxies, and certainly quasars.  We thus conclude that there is a real separation between the different AGN subclasses in the L$_{\rm bol}$/L$_{\rm Edd}$ and L$_{\rm FIR}$ plane.

\begin{table}
\begin{center}
\begin{tabular}{lcc}
\multicolumn{3}{l}{{\bf Table 6: Spearman Rank Coefficients}} \\ 
\hline

\multicolumn{1}{c}{AGN type} & \multicolumn{1}{c}{ r$_{\rm S}$ } & P$_{\rm S}$   \\

\multicolumn{1}{c}{(1)} & \multicolumn{1}{c}{(2)} & (3) \\ 
\hline\hline
ALL & 0.65 & 6.9$\times$10$^{-17}$\\
LINERs & 0.62 & 7.8$\times$10$^{-5}$\\
Seyferts & 0.49 & 2.0$\times$10$^{-4}$\\
RQQ & 0.19 & 0.50\\
RL AGN & 0.70 & 5.2$\times$10$^{-3}$\\
NLS1s & 0.31 & 0.27\\
\hline\hline
\end{tabular}
\end{center}
\tablecomments{\scriptsize{\bf Columns Explanation:} Col(1):AGN subclass; Col(2):  Spearman rank correlation coefficient between L$_{\rm bol}$/L$_{\rm Edd}$ and L$_{\rm FIR}$ for each AGN subtype; Col(3): Probability that the correlation would occur by chance. }
\end{table}

\subsection{{\it ISOPHOT-S} PAH Feature Luminosities and L$_{\rm PAH}$ /L$_{\rm FIR}$ Ratios }

In Table 7, we list the 6.2 $\mu$m  emission feature fluxes and upper limits for all objects in the expanded AGN sample with archival or previously published {\it ISO} observations.  Of the 21 observations we reduced in this work, we report only 4 firm detections.  Combined with published fluxes, there are a total of 28 firm detections of the 6.2 $\mu$m emission feature in our expanded AGN sample.

\begin{table*}
\begin{center}
\begin{tabular}{lccccc}
\multicolumn{6}{l}{{\bf Properties of the {\it ISO} sample }} \\
\hline\hline
\multicolumn{1}{c}{Galaxy} & Distance & Optical & Hubble & log & 
F$_{PAH-6.2}$\\
\multicolumn{1}{c}{Name} & (Mpc) & Class & Type & L$_{FIR}$& 
$\times$10$^{-15}$\\
\multicolumn{1}{c}{(1)} & (2) & (3) & (4) & (5) & (6)\\
& & & & & \\
\hline
\hline
NGC3031 & 4 & LINER/S1.5 & SA(s)ab & 8.4 & $<$6.03 \\
NGC5194 & 8 & S2/LINER & SA(s)bc;pec & 9.8 & 2.20\,$\pm$\,0.18 \\
NGC4486 & 17 & L2 & E+0-1;pec & 8.3 & $<$1.71  \\
NGC4579& 17 &  L1.9 & SAB(rs)b & 9.5 & 1.47\,$\pm$\,0.16 \\
NGC4374 & 18 & L2 & E1 & 8.5 & $<$1.50 \\
IC 1459 & 23 & LINER & E3 & 8.6 & $<$1.78 \\
NGC6240 & 98 & LINER & I0:;pec & 11.3 & 4.52 \\
MRK273 & 151 & LINER & Ring galaxy & 11.8 & 2.14\tablenotemark{a} \\
NGC4151 & 9 & Sy1.5 & SAB(rs)bc & 9.1 & $<$5.91 \\
NGC3227 & 16 & Sy1 & SAB(s)pec & 9.6 & 4.31\,$\pm$\,0.24\tablenotemark{b} \\
NGC3982 & 16 & Sy1 & SAB(r)b & 9.6 & 2.55\,$\pm$\,0.25\tablenotemark{b} \\
NGC1566 & 20 & Sy1 & SAB(rs)bc & 10.1 &  3.70\,$\pm$\,0.49\tablenotemark{b} \\
NGC3516 & 36 & Sy1 & SB(s)0: & 9.5 & 1.44\,$\pm$\,0.24\tablenotemark{b} \\
NGC4593 & 36 & Sy1 & SB(rs)b  & 9.8 & 0.82\,$\pm$\,0.16\tablenotemark{b} \\
NGC3783 & 39 & Sy1.5 & (R')SB(r)a & 9.9 & $<$0.71 \\
IC 4329A & 65 & Sy1.2 & SA0+: sp & 10.0 & 8.87\,$\pm$\,0.26\tablenotemark{b} \\
NGC 7469 & 66 & Sy1.2 & (R')SAB(rs)a & 10.0 & 4.45\,$\pm$\,0.40 \\
NGC5548 & 68 & Sy1 & SA(s)0/a & 9.9 & $<$0.51\tablenotemark{b} \\
Mrk 79 & 89 & Sy1.2 & SBb & 9.9 & $<$0.71 \\
Mrk 590 & 106 & Sy1.2 & SA(s)a: & 10.1 & 1.11\,$\pm$\,0.12\tablenotemark{b} \\
Mrk 817 & 127 & Sy1.5 & SBc & 10.7 & 1.45\,$\pm$\,0.16\tablenotemark{b} \\
Ark 120 & 130 & Sy1 & Sb/pec & 10.2 & 1.23\,$\pm$\,0.16\tablenotemark{b} \\
Mrk 509 & 139 & Sy1.2 & Compact & 10.6 & 0.63\,$\pm$\,0.08\tablenotemark{b} \\
NGC1386 & 12 & Sy2 & SB(s)0+ & 9.1 & 2.16\,$\pm$\,0.24\tablenotemark{b}\\
NGC 1068 & 15 & Sy2 & (R)SA(rs)b; & 10.8 & $<$3.44 \\
NGC5273 & 16 & Sy2 & SA(s)0 & 8.6 & 0.71\,$\pm$\,0.11\tablenotemark{b} \\
NGC7213 & 24 & Sy2 LINER & SA(s)0; & 9.6 &  $<$1.77 \\
IC 5063 & 46 & Sy2 & SA(s)0+: & 10.2 &  $<$0.34 \\
Mrk3    & 56 & Sy2 & S0 & 10.3 & 0.64\,$\pm$\,0.11\tablenotemark{b} \\
NGC1667 & 60 & Sy2 & SAB(r)c & 10.7 & 3.56\,$\pm$\,0.16\tablenotemark{b} \\
Mrk 1 & 64 & Sy2 & (R')S? & 10.2 &  $<$0.13 \\
Mrk 533 & 117 & Sy2 & SA(r)bc pec; & 11.1 &
3.34\,$\pm$\,0.23\tablenotemark{b} \\
UGC 6100 & 119 &  Sy2 & Sa? & 10.2 & 0.71\,$\pm$\,0.08\tablenotemark{b} \\
NGC7603 & 120 &  Sy2 & SA(rs)b & 10.4 & 1.60\,$\pm$\,0.17\tablenotemark{b} \\
NGC4051 & 8 & Sy1.5/NLS & SAB(rs)bc & 9.1 &
1.25\,$\pm$\,0.24\tablenotemark{b} \\
Mrk 766 & 52 & Sy1.5/NLS & (R')SB(s)a: & 10.2 & $<$0.67\tablenotemark{b} \\
PG2130+099 & 244 & Sy1/RQQ & (R)Sa    & 10.6 & $<$0.62 \\
PG0804+761 & 400 & Sy1/RQQ & $\cdots$ & 10.6 & $<$0.97 \\
PG1613+658 & 516 & Sy1/RQQ & Elliptical & 11.5 & $<$0.30 \\
PG0157+001 & 677 & Sy1/RQQ & Bulge/disc & 12.1 & $<$0.45 \\
PG1700+518 & 1243 & Sy1/RQQ & BALQSO & 11.9 & $<$1.02 \\
3C 390.3 & 227 & Sy1/RLQ &  Opt.var.;BLRG  & 10.2 & $<$0.24 \\
\hline
\end{tabular}
\end{center}
\tablenotetext{a}{fluxes from
Spoon et 
al. (2002)}
\tablenotetext{b}{fluxes from Clavel et al. (2000)}
\tablecomments{\noindent{\scriptsize {\bf Columns Explanation:} Col.(1):
Common Source 
Names;  Col.(2): Distance in Mpc; Col.(3): Optical
Classification/AGN 
Class; Col.(4): Hubble Type; Col.(5): Far-infrared luminosities
(in 
units of solar luminosities: L$\odot$) correspond to the
40-500$\mu$m 
wavelength interval and were calculated using the IRAS 60 and
100 $\mu$m 
fluxes according to the prescription of Sanders and Mirabel
(1996), See 
Table 1 above; Col.(6): Flux of the 6.2$\mu$m PAH feature 
($\times$10$^{-15}$) in units of W m$^{-2}$.  Upper limits
correspond to 
3$\sigma$ values.}}
\end{table*}

The FIR luminosity used to construct Figures 5 and 6 can include or be entirely from thermally reprocessed radiation from the AGN. In this case, the FIR luminosity will increase with AGN power as measured by the Eddington ratio.  Such an increase can induce the correlation between L$_{\rm bol}$/L$_{\rm Edd}$ and L$_{FIR}$ displayed in Figures 5 and 6, invalidating any implied association between the SFR and the mass accretion rate.    Since the PAH feature flux is found to be directly proportional to the FIR flux in normal and starburst galaxies but is absent or weak in galaxies dominated by AGN (e.g., Genzel et al. 1998, Rigopoulou et al. 1999, Clavel et al. 2000), we can use our 6.2 $\mu$m  emission feature fluxes to investigate whether the fraction of L$_{FIR}$ increases with L$_{\rm bol}$/L$_{\rm Edd}$ in our expanded AGN sample.  Figure 9 shows the relationship between L$_{\rm PAH}$ /L$_{\rm FIR}$ and L/L$_{\rm Edd}$.  If the fraction of the FIR luminosity increases with Eddington ratio, then we should see a decrease in L$_{\rm PAH}$ /L$_{\rm FIR}$ with L$_{\rm bol}$/L$_{\rm Edd}$.  Although the data are limited, Figure 9 reveals no correlation, strongly suggesting that the correlation between L$_{\rm bol}$/L$_{\rm Edd}$ and L$_{FIR}$ for this sample of AGN is NOT an artifact of an increasing contribution to the FIR emission from dust heated by an AGN with Eddington ratio. 

\begin{figure*}
{\includegraphics[width=16cm]{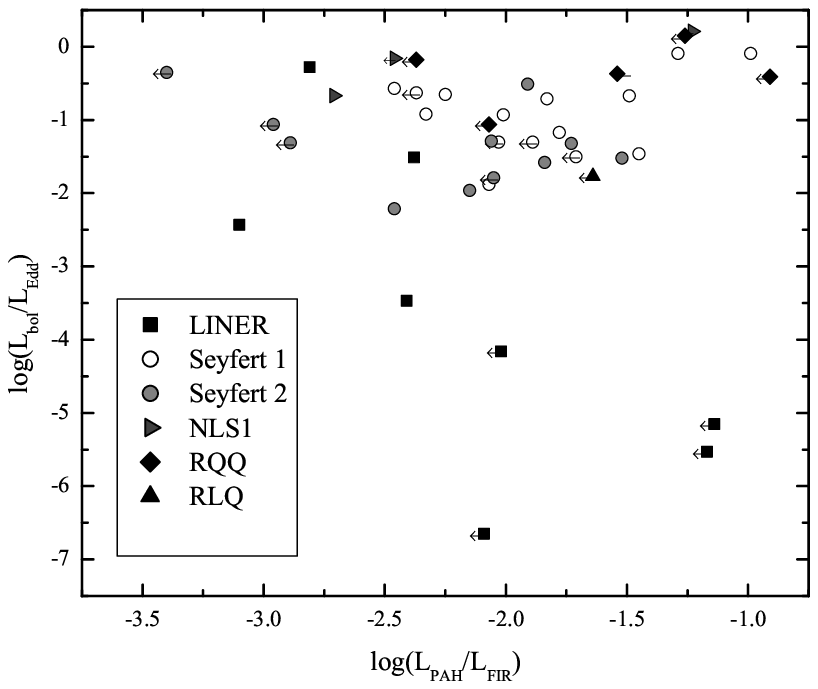}}\\
\caption[]{Plot of the ratio the bolometric luminosity to the Eddington luminosity as a function of the ratio of the 6.2$\mu$m PAH luminosity and L$_{FIR}$ for all of the galaxies in our expanded AGN sample with 6.2$\mu$m data. No correlation is found between these two quantities, implying that the contribution from the AGN to L$_{FIR}$ is indeed minimal.}
\end{figure*}

\section{Discussion}
\subsection{ L$_{FIR}$ as a SFR Indicator}
Since the FIR luminosity is widely used as a direct measure of the SFR in galaxies (Kennicutt 1998; Lehnert \& Heckman 1996; Meurer et al. 1997; Kewley et al. 2002), the correlations presented in D05 and expanded upon in Figures 5 and 6 hint at the possibility of a fundamental link between accretion onto the black hole and the SFR in the host galaxy.  The reliability of using L$_{\rm FIR}$ to estimate the SFR rests on the assumption that dust heating is dominated primarily by the radiation field from young stars and that the dust surrounding these stars absorbs all of the optical/ultraviolet (OUV) radiation and reemits it in the FIR.  While this may be a good assumption in optically thick intensely star forming galaxies, it may not hold for all galaxies in our sample. In converting the FIR luminosity to a SFR in our expanded AGN sample of galaxies, the following questions need to be addressed: 1) What fraction of L$_{\rm FIR}$  arises from dust at large distances from and heated primarily by the AGN? 2) What is the contribution from old stars to L$_{\rm FIR}$? 3) How much of the bolometric luminosity from young stars is emitted in the FIR?

\subsubsection{ What fraction of the FIR luminosity in AGN arises from dust at large distances from and heated primarily by the AGN}
There is general agreement that the AGN radiation field, which heats the dust to higher temperatures than does the starburst, is  responsible for  the near-IR and MIR emission in AGN (e.g., Wilkes et al. 1999; van Bemmel \& Dullemond 2003; Farrah et al. 2003).  The FIR emission, however, can in principle arise from starburst-heated dust as well as cooler dust heated by and located farther from the AGN.  For nearby quasars, several studies claim that the entire near-IR, MIR, and FIR SED can be explained purely by AGN heating (e.g. Sanders et al. 1989, Kuraszkiewicz et al. 2003, Siebenmorgen et al. 2004).  However, these studies do not rule out the possibility that starbursts provide the dominant heating source in powering the FIR.  Indeed, detailed modeling of the SEDs of hyperluminous AGN ($L_{\rm FIR}$ $>$ 10$^{13}$${\rm L_{\odot}}$) suggest that star formation provides the dominant heating source behind their FIR emission (Rowan Robinson 2000), and the lack of correlation between the MIR and FIR of several well-studied PG quasars strongly suggests that the two wavelength regimes are dominated by different heating sources (Haas et al. 1999).  In addition, if AGN heating dominates the cooler FIR-emitting dust, then there should be a correlation between quasar OUV and FIR luminosity which is not seen (e.g., McMahon et al. 1999; Isaak et al. 2002; Priddey et al. 2003). In Seyfert galaxies, the spatially extended distribution of the FIR emission as well as the similarity between their IR SEDs with starbursts suggests that the AGN is not responsible for the bulk of their FIR emission (Espinosa et al. 1987).  The IR properties of LINERs are currently not well-studied but since their central AGN are generally weak, it is unlikely that they provide the dominant heating source behind their FIR emission.  Furthermore the lack of correlation between the 2-10 keV X-ray and FIR fluxes for the AGN-LINERs plotted in Figures 5a and 5b (section 4.2) strongly suggests that FIR is not related to the AGN in our sample of LINERs.  In addition, the relationship between L$_{\rm PAH}$ /L$_{\rm FIR}$ and L$_{\rm bol}$/L$_{\rm Edd}$ discussed in Section 4 argues against AGN heating being the primary mechanism behind the FIR for the galaxies plotted in Figure 9.  If L$_{\rm FIR}$ is dominated by AGN heating, then L$_{\rm PAH}$ /L$_{\rm FIR}$ should decrease with AGN power, as measured by the Eddington ratio.  Although the data are limited, this may be the most definitive discriminator between the relative contributions of starbursts and AGN to the FIR emission in AGN.  We make the explicit assumption in this work that AGN heating of the dust is not responsible for the bulk of the FIR emission for all galaxies in our expanded AGN sample and explore the consequences.  More extensive studies of the PAH feature in AGN with Spitzer can help confirm this assumption.

\subsubsection{ What is the contribution from old stars to L$_{\rm FIR}$ and is the dust opacity high in our sample?}
A significant fraction of the galaxies in our expanded AGN sample are bulge-dominated. Some studies indicate that in early-type galaxies, the general stellar radiation field may make a significant contribution to the dust heating and that the dust opacity may be small (e.g., Sauvage \& Thuan 1994; Mazzei \& de Zotti 1994).  Contrary to these findings, a detailed comparison of the H$\alpha$ and FIR emission in a large sample of normal galaxies suggests that the FIR emission {\it can} be associated primarily with star formation and that the widely adopted SFR- L$_{\rm FIR}$ calibration ratio usually applied to starbursts from Kennicutt (1998) can be applied in all Hubble types, including early-type spirals (Kewley et al. 2002).  The ambiguity in the contribution to the dust heating by old stars affects the calibration of the SFR in terms of L$_{\rm FIR}$.  Since our expanded AGN sample was inhomogenously constructed and spans a large range in luminosity and Hubble types, it may not be appropriate to apply the Kennicutt (1998) SFR- L$_{\rm FIR}$ calibration derived for starbursts for all objects in the sample.  Instead, we chose to adopt the calibration ratio from the recent work by Bell (2003).  Using a large and diverse sample of normal and starburst galaxies and multiple SFR indicators, they find that the old underlying stellar population's contribution to L$_{\rm FIR}$ increases as the galaxy's luminosity decreases.   Using their calibration ratio, the host galaxy's SFR can be calculated:

\begin{equation}
 \label{belfir}
{\rm SFR_{FIR}}(M_{\odot}{\rm yr^{-1}})     \frac{1.75 f L_{\rm FIR}}{6.38\times10^{9}{\rm L_{\odot}}}
       =\frac{f L_{\rm FIR}}{3.63\times10^{9}{\rm L_{\odot}}}, 
\end{equation} 
where 
\begin{equation} 
\label{firf} f = \left\{ \begin{array}{ll}
 1 + \sqrt{5.71\times10^{9}\,{\rm L_{\odot}}/L_{\rm FIR}} & L_{\rm FIR} > L_c \\
 0.75(1 + \sqrt{5.71\times10^{9}{\rm L_{\odot}}/L_{\rm FIR}}) & L_{\rm FIR} \le L_ c,
\end{array} \right.
\end{equation}
and $L_c=5.71\times10^{10}$L$_{\odot}$. $L_{\rm FIR}$ is the luminosity corresponding to the FIR flux as defined above, and the factor of
1.75 in Equation 6 converts this to a luminosity representative of the full ($8-1000\,\mu$m) mid- to far-infrared spectrum (for details see Bell 2003).

The resulting SFR agrees to within a factor of 2 of the SFR derived using the Kennicutt (1998) calibration for all galaxies in our expanded AGN sample with the exception of only two LINERs with elliptical hosts, where slightly more discrepant factors are seen ($\sim$ 3 and 6).  Bell points out that agreement with the Kennicutt (1998) value arises because of the competing effects of the contribution of old stars, which reduces the derived SFR, and the reduction in dust opacity with decreasing luminosity, which increases the derived SFR.  In calculating SFRs, we stress that we have made the explicit assumption in this work that AGN heating plays an insignificant role in the FIR emission in our expanded AGN sample.

\subsection{ The SFR-Accretion Rate Connection}
The discovery of the correlation between black hole mass and stellar velocity dispersion (Gebhardt et al. 2000, Ferrarese \& Merritt 2000) has spawned numerous speculations on the connection between the growth of black holes and the formation of galactic bulges. A number of theoretical models have attempted to explain the relationship, invoking radiative or mechanical feedback from the black hole on the gas supply in the bulge (e.g., Silk \& Rees 1998; Haehnelt, Natarajan, \& Rees 1998; Blandford 1999; King 2003; Wyithe \& Loeb 2003), merger-driven starbursts with black hole accretion (e.g. Haehnelt \& Kauffman 2000), and stellar captures by the accretion disk feeding the hole (e.g., Zhao, Haehnelt \& Rees 2002).  However, the case remains ambiguous which one is correct or, for that matter, whether there really is a causal connection between the birth and growth of black holes and the formation and evolution of galaxies.   Most previous studies have focused on expanding the number of black hole and bulge mass estimates in the various AGN subclasses in an attempt to reconstruct the accretion and star formation history in galaxies during various phases of accretion activity.  If black holes grow primarily when they are accreting--i.e., when they are AGN, a complementary and more direct constraint to theoretical models is to determine the relationship between the mass accretion rate and the spheroidal star formation rate in the various manifestations of nuclear activity in galaxies.  If there is a constant ratio between the accretion rate and the star formation rate (SFR) associated with the bulge, with a proportionality constant independent of time, galaxy type, merger status, and accretion activity, this can have a tremendous impact on our understanding of galaxy formation and evolution.

Using the SFR calibration ratio described above, together with an assumption of the radiative efficiency of accretion, we can convert Figure 6 to a plot of SFR vs. mass accretion rate, $\dot{M}$ in our sample of galaxies.  Making the first order assumption that all AGN in our sample have a radiative efficiency $\eta$ =0.1, the standard factor used for a geometrically thin, optically thick accretion disk (Shukura \& Sunyaev 1973, Narayan \& Yi 1995), we plot in Figure 10 the SFR vs. $\dot{M}$ for our expanded AGN sample.  A linear fit yields the following relationships for the AGN-LINER class:

\begin{equation}
\log {\rm SFR}=(0.45\pm0.06)\log \dot{M} +(1.46\pm0.25),
\end{equation}
where ${\rm SFR}$ and $\dot{M}$ are both in units of M$_{\odot}$${\rm yr^{-1}}$.

From Figure 10, there is a clear distinction in slope and intercept between the various AGN subclasses.  The regression line for the Seyfert class is:
\begin{equation}
\log {\rm SFR}=(0.89\pm0.07)\log \dot{M} +(1.68\pm0.12),
\end{equation}

In Section 4.3, we showed that the apparent separation between the various AGN subclasses in the L$_{\rm bol}$/L$_{\rm Edd}$ and L$_{\rm FIR}$ plane, which gives rise to the separation in the SFR and $\dot{M}$ plane seen in Figure 10, is not likely to be due to systematic effects in estimating $ L_{\rm bol}$, $L_{\rm Edd}$, or $L_{\rm FIR}$ in our expanded AGN sample.  Figure 11 shows the distribution of the SFR/$\dot{M}$ ratio for the sample.  NLS1s and RQQ, generally considered to be characterized by very high accretion rates, generally have the smallest SFR/$\dot{M}$ values, followed by Seyferts, RL AGN, and finally, LINERs.   If we make the crude assumption that the FIR luminosity is at least loosely correlated with star formation in the bulge, at face value, Figures 10 and 11 imply that the growth of the black hole by accretion does not always match the growth of the bulge during all phases of galaxy evolution. Is there prodigious black hole growth without major star formation in the most highly accreting local sources compared with the weakly accreting sources?  

\begin{figure*}[]
{\includegraphics[width=16cm]{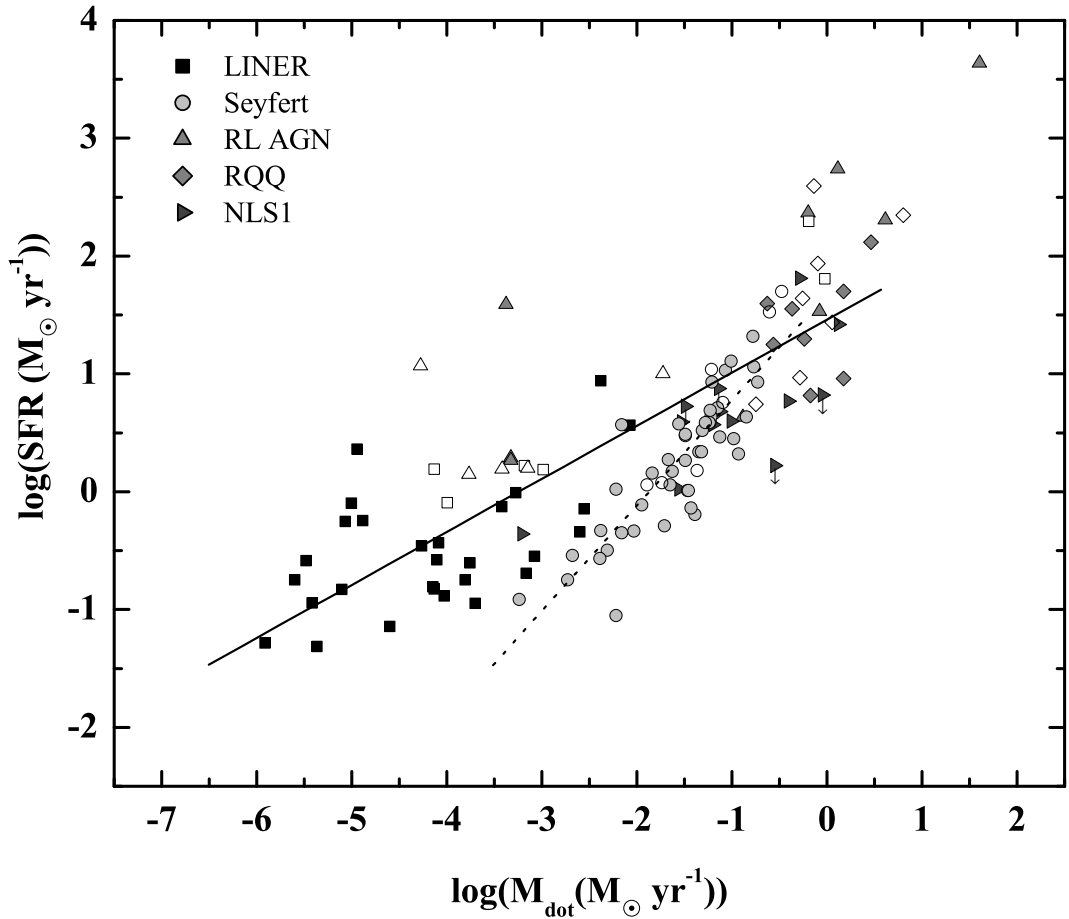}}\\
\caption[]{Log plot of the Star Formation rate in $M_{\odot}$${\rm yr_{-1}}$ as a function of the Mass Accretion Rate in $M_{\odot}$${\rm yr_{-1}}$ for the `129 AGN in our expanded AGN sample with black hole estimates. The dotted line represents the linear fit to the Seyfert data and the solid line represents the linear fit to the LINER data.  AGN that belong to merging or interacting systems are indicated by the open symbols.}
\end{figure*}

\begin{figure}[]
\epsscale{.9}
{\includegraphics[width=9cm]{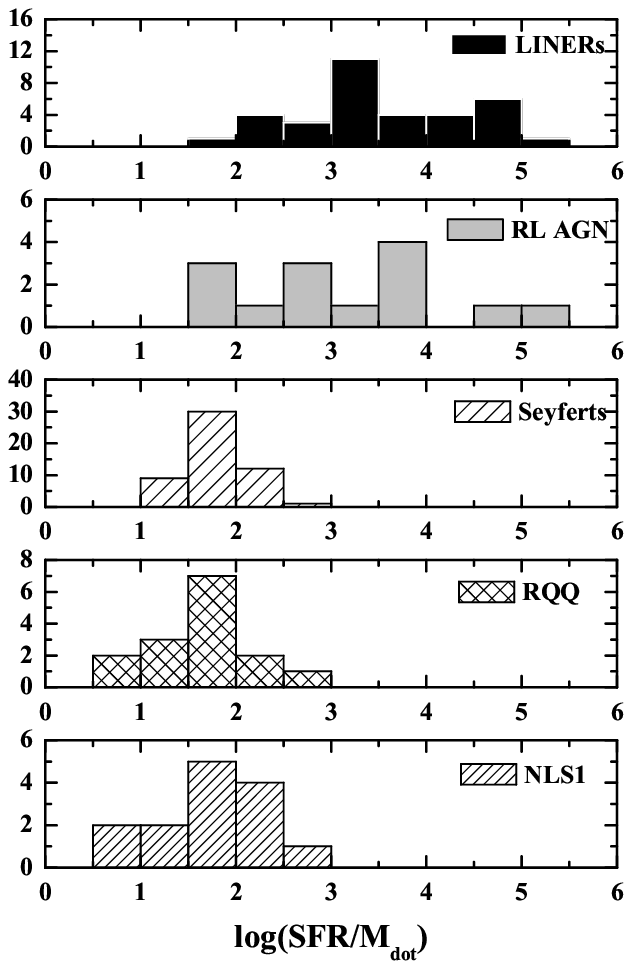}}\\
\caption[]{Histogram of the ratio of the star formation rate to the mass accretion rate for the 129 AGN in this work that have black hole estimates.}
\end{figure}

Alternatively, the separation seen in Figures 10 and 11 could simply be an artificial consequence of our assumption that $\eta$ is the same for all sources, while in fact the radiative efficiency for the objects at low $\dot{M}$  is likely to be much smaller (and possibly a decreasing function of $\dot{M}$; Narayan \& Yi 1995).  Such an effect would alter both the slope and intercepts of the regression lines displayed in Figure 10 as well as introduce greater dispersion in the RL AGN and LINERs which span a greater range in L$_{\rm bol}$/L$_{\rm Edd}$ consistent with what is seen in Figures 10 and 11.  Indeed, a transition between a radiatively efficient, geometrically thin, optically thick accretion flow, and a radiatively inefficient, geometrically thick and optically thin flow is theoretically expected to occur at low values of $\dot{M}$  (Rees et al. 1982; Narayan \& Yi 1995). Applying a single value for $\eta$ to a sample of AGN (as is commonly done in the literature), particulary in our sample where L$_{\rm bol}$/L$_{\rm Edd}$ varies widely, is therefore likely to be incorrect.   Most of the LINERs and many of the RL AGN in our expanded AGN sample (see Tables 2 and 4) have low L$_{\rm bol}$/L$_{\rm Edd}$ values.  If the accretion flow in those objects with low Eddington ratios is characterized by systematically lower $\eta$ values compared with the rest of the sample, the separations seen in Figures 10 and 11 can be removed.  If we adopt $\eta$ = 0.001 for LINERs and the RL AGN with low values of L/L$_{\rm Edd}$, Figures 10 and 11 show that the SFR/$\dot{M}$ ratio becomes more consistent amongst the various AGN types.  This value of $\eta$ is consistent with radiatively inefficient accretion models (RIAF; Quataert 2003) and is in fact compatible with the values of $\eta$ calculated directly based on the Bondi accretion rates and a detailed spatial analysis of a sample of radio galaxies with low Eddington ratios (Donato, Sambruna, \& Gliozzi 2005).  In order to truly understand the relationship between the mass accretion rate and the host galaxy SFR in the various AGN subclasses, it is necessary to determine the radiative efficiencies and decouple their effects on the derived mass accretion rates for the entire sample. However, our limited dataset and theoretical uncertainties do not allow us to expand further on this point, or to give any robust indication on the true value of $\eta$ for all galaxies in our sample.

If there is a variation in the SFR/$\dot{M}$ ratio as function of AGN type and activity level, this may have an important consequence on our understanding of galaxy evolution and black hole growth.  Using our expanded AGN sample, we can also investigate whether there are any systematic trends in the SFR/$\dot{M}$ ratio as a function of interaction status.  Our sample includes 22 AGN that are in either merging or interacting pairs. Several models assume black hole growth matches bulge growth exactly during a merger with subsequent growth of the bulge being regulated by AGN feedback (e.g. Haehnelt \& Kauffman 2000). If this scenario holds, one might expect to see variations in the SFR/$\dot{M}$ ratio as a function of AGN type, activity level, and merger status.  Indeed, recent hydrodynamical simulations that simultaneously follow star formation and the growth of black holes during galaxy-galaxy collisions shows that SFR/$\dot{M}$ varies with time as energy released by the AGN expels enough gas to quench both star formation and further black hole growth (Di Matteo, Springel, \& Hernquist 2005).  In Figure 10, merging or interacting galaxies are indicated with open symbols.  Apart from the fact that the merging or interacting galaxies are concentrated at high values of $\dot{M}$, there is no apparent distinction in their SFR vs. $\dot{M}$ relation.  Clearly there are not enough data to make definitive conclusions.  A more extensive analysis that includes a larger population of mergers would be required to address this important question.

There have been a few recent studies on the connection between the SFR and mass accretion rate in AGN.  Using the SDSS observations of 123,000 low redshift galaxies, Heckman et al. (2004) find that the global volume-averaged SFR/$\dot{M}$ ratio  is approximately 1000 in bulge-dominated systems, in agreement with ratio of bulge to black hole mass implied by the M$_{\rm BH}$ vs. $\sigma$ relationship (Marconi \& Hunt 2003).  This ratio is significantly higher than the the SFR/$\dot{M}$ ratio for the majority of galaxies in our sample, which target solely definitive AGN (see Figure 11).  For example, the average SFR/$\dot{M}$ ratio for the Seyfert and RQQ subclasses is approximately 100 - a value that can be compatible with the global ratio of 1000 if there is a duty cycle for the AGN of about 10\%. Interestingly, Hao et al. (2005) recently found a significantly higher SFR/$\dot{M}$ ratio ($\sim$500) in a small sample of infrared-selected QSOs compared with the majority of AGN in our expanded AGN sample.  Since this ratio is determined in IR-bright QSOs, there is potentially enhanced star formation in these sources compared with the standard RQQ included in our sample.

We note that our estimate of the SFR is based on the total L$_{\rm FIR}$ from the host galaxy.  As we have pointed out, the FIR emission can include an AGN contribution, a contribution from old stars, and probably, most importantly, star formation taking place in the disk, all of which would result in an overestimate to the bulge SFR. Indeed, in spiral hosts, the total FIR emission may be dominated by star formation in the disk (e.g., Fukugita et al. 1998; Benson et al. 2002; Hogg et al. 2002). Spatially resolved observations of the PAH emission in these galaxies - possibly the most robust indicators of the SFR in AGN - can potentially provide a more accurate determination of bulge-dominated star formation.  Future spectroscopic imaging observations of these features with {\it Spitzer} in a statistically significant sample of nearby AGN can potentially provide important advances in our understanding of the observed correlation of the black hole and bulge mass and its relationship to the birth, growth, and evolution of black holes and galaxies. 

\section{Conclusion}

We have studied the relationship between the Eddington ratio, L/L$_{\rm Edd}$, as a function of the FIR luminosity, L$_{\rm FIR}$, of the host galaxy, found to be correlated over seven orders of magniture in L$_{\rm bol}$/L$_{\rm Edd}$ in our previous work (Paper II), in a sample of 34 LINERs with confirmed hard X-ray nuclear point sources.  This sample builds on our previously published proprietary and archival X-ray observations from {\it Chandra} by including the remaining 25 LINERs in the {\it Chandra} archive for which black hole masses and FIR luminosities have been previously published.   We combined our sample with a larger sample of AGN with reliable black hole masses and bolometric luminosities drawn from the literature.  The entire sample discussed in this work consists of 129 confirmed AGN: 34 AGN-LINERs, 52 Seyferts, 14 radio-loud AGN, 15 QSOs, and 14 narrow line Seyfert 1s.  Our main results are as follows.  

\begin{enumerate}
\item Of the 25 LINERs presented in this article, 13 show compact hard nuclear cores coincident with the radio or {\it 2MASS} nucleus, with a luminosity L$_X$ (2-10 keV) $\geq$ 2 $\times$ 10$^{38}$ ergs s$^{-1}$.   The nuclear 2-10 keV luminosities for the 25 galaxies range from $\sim$ 2 $\times$ 10$^{38}$ ergs s$^{-1}$ to $\sim$ 1 $\times$ 10$^{42}$ ergs s$^{-1}$.  

 \item Combining these observations with our previously published work, we find that 50\% (41/82) of LINERs have hard nuclear X-ray cores consistent with an AGN.

\item We find a significant correlation between the Eddington ratio as a function of  L$_{\rm FIR}$ that extends over almost nine orders of magnitude in L$_{\rm bol}$/L$_{\rm Edd}$. Using archival and previously published observations of the 6.2 $\mu$m PAH feature, we find that it is unlikely that dust heating by the AGN dominates the FIR luminosity in our sample of AGN.  Our results may therefore imply a fundamental link between the mass accretion rate ($\dot{M}$), as measured by the Eddington ratio, and star formation rate (SFR), as measured by the FIR luminosity. 

\item Apart from the overall correlation, we find that the different AGN subclasses occupy distinct regions in the L$_{\rm FIR}$  and L$_{\rm bol}$/L$_{\rm Edd}$ plane.  This may imply a variation in the SFR/$\dot{M}$ ratio as function of AGN type and activity level. Although data are limited, there seems to be no systematic difference in the derived SFR/$\dot{M}$ ratio in AGN that belong to merging or interacting pairs and those that do not.

 \item Assuming that the radiative efficieny for accretion is 10\% for all AGN in the sample and that the FIR luminosity traces arises solely from dust heated by young stars, we determined the relationship between the mass accretion rate and the host galaxy's SFR.  The average ratios of the SFR/$\dot{M}$ ratio for the Seyfert and RQQ subclasses are approximately 100 - a value that can be compatible with the global volume-averaged ratio of 1000 found in a large sample of bulge-dominated systems observed in the SDSS, if there is a duty cycle for the AGN of about 10\%.

\end{enumerate}

We are very thankful to Davide Donato, Alex Rinn \& Rita Sambruna for their invaluable help in the data analysis, and for numerous instructive discussions. Also we thank Joe Weingartner and Jackie Fischer, for their enlightening and thoughtful comments.  We are also grateful for the very thorough and detailed review by the referee, which helped improve this work.  This research has made use of the NASA/IPAC Extragalactic Database (NED) which is operated by the Jet Propulsion Laboratory, California Institute of Technology, under contract with the National Aeronautics and Space Administration.  SS gratefully acknowledges financial support from NASA grant NAG5-11432 and NAG03-4134X.

\end{document}